\documentstyle[aps,prb,preprint]{revtex}
\begin{document}
\title
{Resonant two-magnon Raman scattering in parent compounds of
high-T$_c$ superconductors.}
\author{Andrey V. Chubukov}
\address
{Department of Physics, University of Wisconsin-Madison,
1150 University ave., Madison, WI 53706\\
and P.L. Kapitza Institute for Physical Problems, Moscow, Russia}
\author{David M. Frenkel}
\address
{Texas Center for Superconductivity,\cite{byline}
University of Houston, Houston,
TX 77204-5932\\
and
Department of Physics, Science and Technology Center
for Superconductivity,\\
University of Illinois at Urbana-Champaign,
1110 West Green Street, Urbana, IL 61801}
\date{today}
\maketitle
\begin{abstract}
We propose a theory of two-magnon Raman scattering from the insulating
parent compounds of high-T$_c$ superconductors, which contains
information not only on magnetism, but also on the electronic properties
in these materials.
We use spin density wave formalism for the Hubbard model, and study
diagrammatically the profile of the two-magnon scattering and its
intensity dependence on the incoming photon frequency $\omega_i$ both
for $\omega_i \ll U$ and in the resonant regime, in which the energy of the
incident photon is close to the gap between conduction and valence bands.
In the nonresonant case, we identify the diagrams which contribute to the
conventional Loudon-Fleury Hamiltonian.
In the resonant regime, where most of the experiments have been done,
we find that the dominant contribution to Raman intensity comes from
a different
diagram, one which allows for a simultaneous vanishing of all three of its
denominators (i.e., a triple resonance).
We study this diagram in detail and show
that the triple resonance, combined with the spin-density-wave dispersion
relation for the carriers, explains the unusual features found in the
two-magnon profile and
in
the two-magnon peak intensity dependence on the incoming photon frequency.
In particular, our theory predicts a maximum of the two-magnon peak
intensity right at the upper edge of the features in the optical data,
which has been one of the key experimental puzzles.
\end{abstract}
\narrowtext

\section{Introduction}

There is a widespread belief that strong electron-electron
correlations in the high-T$_c$ compounds may hold a clue to
the phenomenon of high-temperature
superconductivity.\cite{anderson,andersonschrieffer,pinesanderson}
One of the manifestations of these correlations is in the fact
that the insulating parent compounds are antiferromagnets.
An important probe of antiferromagnetism is magnetic Raman
scattering.\cite{oldmerlin,cottamlockwoodbook,singhreview,kleincooperreview}
Its prominent signature in the underdoped high-$T_c$ materials
is a two-magnon peak  observed at about $3000~cm^{-1}$. To first
approximation, this peak can be attributed to inelastic scattering
from  the two-magnon excitations.\cite{fleury}
In fact, Raman experiments yielded the first estimate of the exchange
interaction constant in $La_{2}CuO_{4}$.\cite{fleury,lyons,%
oldjapaneseresonantdata,askandrey1}
The two-magnon peak has also been observed in the electron-doped
materials\cite{tomenoetal} and more recently\cite{askandrey2}
in the underdoped $YBa_2Cu_3O_{6+x}$ materials (up to $x =0.9$).

The traditional framework for understanding the two-magnon Raman scattering
in antiferromagnets has been an effective Hamiltonian for the interaction
of light with spin degrees of freedom known as the Loudon-Fleury
Hamiltonian\cite{loudonfleury}
\begin{equation}
    H=\alpha\sum_{<ij>}
    ({\bf\hat e}_i\cdot{\bf R}_{ij})
    ({\bf\hat e}_f\cdot{\bf R}_{ij})
    {\bf S}_i \cdot {\bf S}_j.
  \label{eq1}
\end{equation}
Here ${\bf\hat e}_i$ and ${\bf\hat e}_f$ are the polarization vectors
of the in- and outgoing photons, $\alpha$ is the (generally poorly known)
coupling constant, and ${\bf R}_{ij}$ is a vector along the bond connecting
two nearest neighbor sites $i$ and $j$.
This equation has been reexamined in recent years to better account for a
possible importance of quantum fluctuations, the four-magnon processes,
and the further neighbor
terms.\cite{singhreview,singhetal,canaligirvin}

The Loudon-Fleury theory is expected to work well when the frequencies of
the incoming
and outgoing photons are considerably smaller than the gap between conduction
and valence bands.
The experimental reality in high-$T_c$ materials is such, however, that the
two-magnon scattering  is observed mostly in or near the so-called
resonant regime, when the frequencies of the ingoing and/or outgoing
photons are close to the gap value.
This is simply a consequence of the fact that the experiments are mostly
done with the visible or near ultraviolet lasers, while the
gap and the upper Hubbard band states lie in the same range of frequences
(roughly between $2~eV$ and $3~eV$) in the high-$T_c$ materials.
The experimental cross-sections vary strongly in this range of incident
photon frequencies,\cite{oldjapaneseresonantdata,oldjapanesedata2,ranliu}
and it becomes an issue whether the Loudon-Fleury Hamiltonian is
still applicable.
Whatever the answer to this question, we still need to have the means
of evaluating the variation of the overall scale of the coupling constant
$\alpha$ in Eq.~(\ref{eq1}) with the incident photon frequency $\omega_i$.

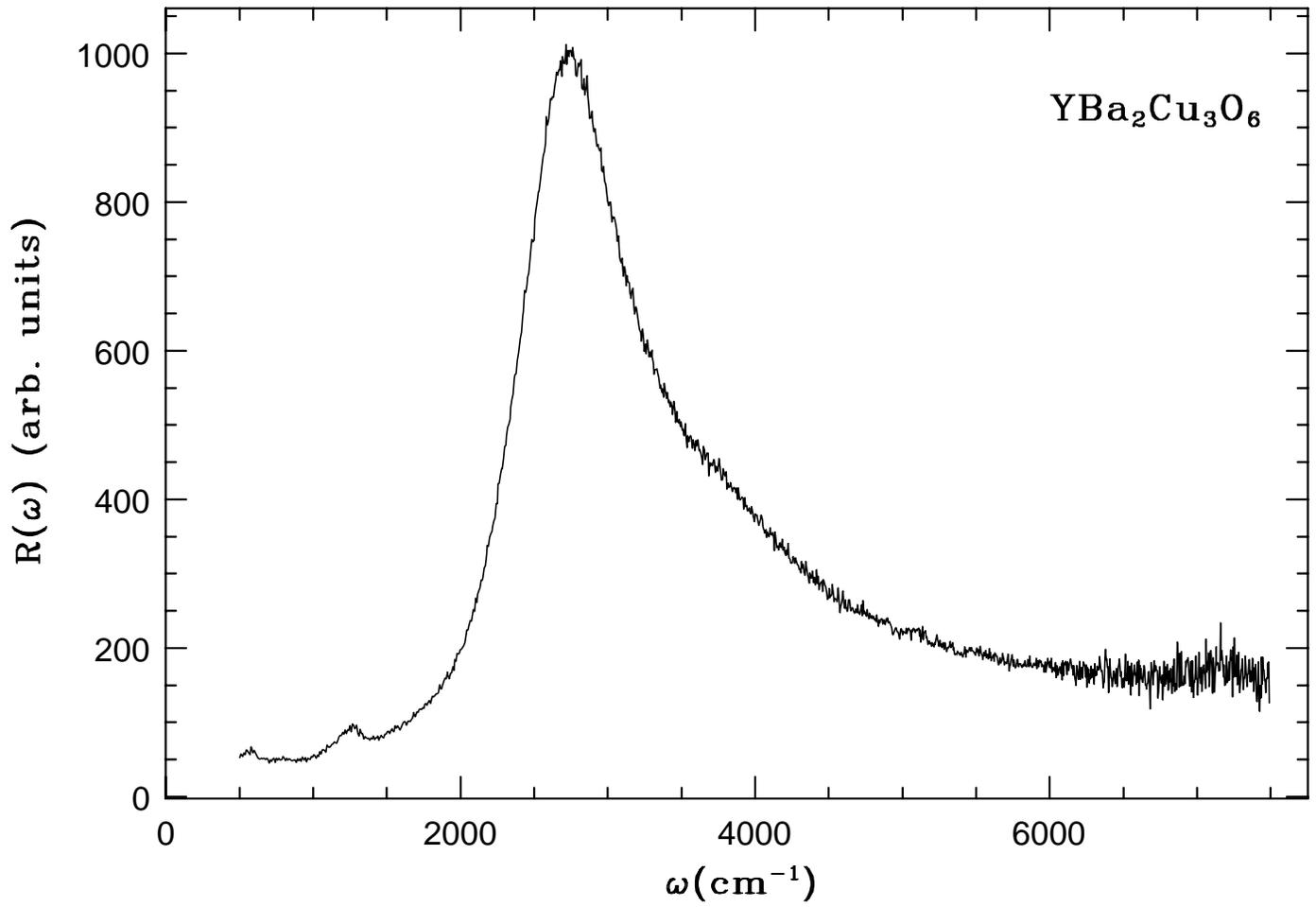
\begin{figure}
\caption{A typical Raman cross-section as a function of transferred photon
frequency. A two-magnon peak is clearly seen. Data courtesy
of the authors of Ref.~\protect\onlinecite{ranliu}.}
\label{twom}
\end{figure}

The profile of the Raman cross-section as a function of the
{\it transferred\/}
photon frequency is shown in Fig.~\ref{twom}.
The two-magnon peak is clearly identifiable.
The behavior of the two-magnon peak height as a function of the {\it
incident\/} photon frequency is presented in Fig.~\ref{peak}, where the
absorptive part of the dielectric constant is also shown for comparison.

The key experimental features that require explanation are the
following:

In Fig.~\ref{twom}:

\noindent
(a) {\it Asymmetry of the two-magnon peak profile\/}: the two-magnon peak is
asymmetric, with the spectral weight shifted to higher frequencies.

\noindent
(b) {\it Selection rules\/}: Loudon-Fleury Hamiltonian predicts no
scattering in the $A_{1g}$ configuration; experimentally in the
resonant regime the $A_{1g}$ cross-section is about
half of that in the $B_{1g}$ geometry.

\noindent
(c) {\it Stability of the two-magnon peak profile\/}: as the incident
photon frequency changes over the
frequency range from $15,000$ to $22,000~cm^{-1}$,
 the two-magnon scattering intensity profile merely
scales with incident frequency, without a noticeable change in shape.
(However, at higher photon energies,
a distortion of the Raman spectrum occurs.)

In Fig.~\ref{peak}:

\noindent
(a) {\it A single peak\/}: ordinarily one might expect two peaks, the so
called ingoing and outgoing
resonances,\cite{cardonareview,martinfalikovreview} and here only one
is observed.

\noindent
(b) {\it Peak location\/}: a comparison with the dielectric constant
shows that the strength of the two-magnon Raman scattering is at its
maximum away from the band edge, in fact right at the upper end of the
features in the optical data that can be interpreted as the particle-hole
excitations between the lower and upper Hubbard bands.
(Phonons, on the contrary, are known to resonate with the
features in the dielectric constant.\cite{heyenetal})

Raman scattering in a one-band model of correlated electrons has been
considered several times in the
literature.\cite{canaligirvin,shastryshraiman1,kampfbrenig}
Shastry and Shraiman\cite{shastryshraiman1} have recently
given a derivation of the Loudon-Fleury Hamiltonian starting from the usual
large-$U$ Hubbard model.
Working in a localized basis, they performed a hopping expansion
controlled by $t/(U-\omega)$, where $t$ and $U$ are the nearest-neighbor
hopping and on-site Coulomb repulsion, and $\omega$ is of the order
of the photon frequencies.
The leading term in the expansion turned out to be the Loudon-Fleury
Hamiltonian, Eq.~(\ref{eq1}).

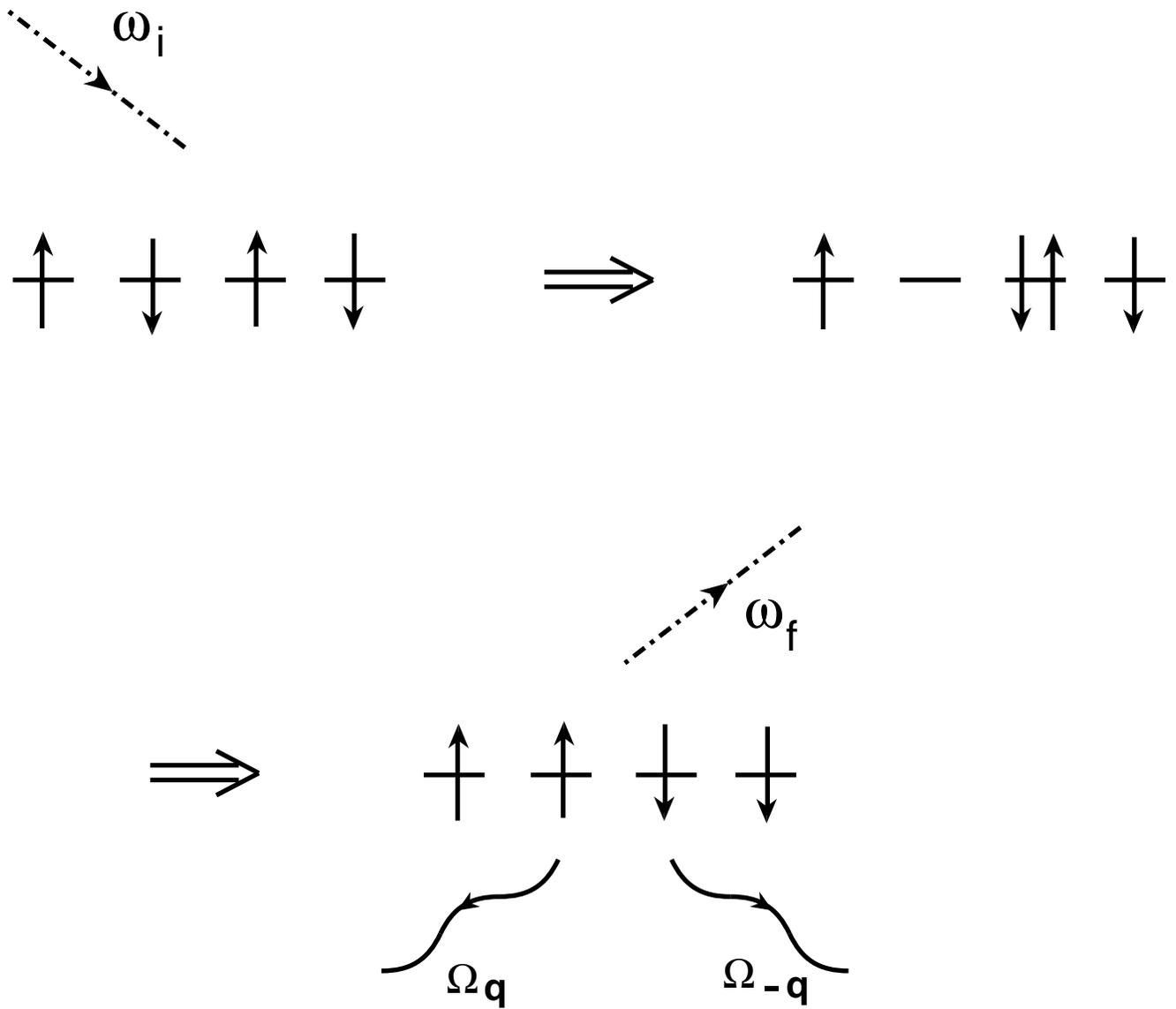
\begin{figure}
\caption{The real-space picture of the two-magnon scattering. Dashed
lines denote photons, and wavy lines denote spin waves.
The explanations are given in the text.
This picture is valid away from resonance.}
\label{shastshrproc}
\end{figure}

A simple, though somewhat crude interpretation of this outcome is to
envision spins in a classical Neel state.
The incoming photon is absorbed by moving, say, a down spin to the
neighboring site occupied by an up spin.
The up spin then returns to the formerly down site, emitting the
outgoing photon.
The overall result is light scattering accompanied by the nearest
neighbor spin exchange, which is encoded in Eq.~(\ref{eq1}).
This process is shown in Fig.~\ref{shastshrproc}.

Notice that the intermediate state in this process has a doubly occupied
site, and is thus of order $U$ above the ground state in energy.
When the photon energy is far smaller than $U$, no further precision
in specifying the intermediate state's energy is needed, and the
leading order term is sufficient.
When $(U-\omega)$ becomes of order $t$ (resonant regime), all
of the terms in the expansion in $t/(U-\omega)$ are of the same order, and
the leading term can no longer provide a solution.
However, it is in this regime that the cross-secions sensitively
depend not just on the magnetic, but also on the carrier properties,
and this makes it a especially important to the field of high-T$_c$
superconductivity.

We will discuss the form of the Raman vertex in both the nonresonant
and  resonant regimes later in the paper and here merely note that, on
physical grounds, the multiple hops affect the single-particle Green's
function in two distinct ways.
First, the multiple hops is a way to generate the band structures of the
doubly occupied sites and holes, i.e., they produce a dispersion
in the coherent part of the single-particle Green's function.
Second, they produce a distortion of the spin background around each
hole which gives rise to the incoherent part of the Green's function.
Since resonance is a phenomenon contingent upon the vanishing of the
denominators in perturbation theory,
 it is plausible to assume that the incoherent parts
of the Green's functions are of lesser importance.
We thus model our particle and hole Green's functions as fully coherent,
but having a certain dispersion, which at half-filling we take to
be that of the spin density wave (SDW) solution of the Hubbard
model.\cite{schriefferwenzhang}
We regard our ability to address the experimental results on the basis
of this dispersion relation as partial evidence of the validity of the
SDW picture not only for the antiferromagnetism, but also for
the hole dynamics.

We wish to stress that this is not a trivial distinction.
There have been many interesting theoretical studies of the single
hole problem in the Hubbard and the t-J
models.\cite{schriefferwenzhang,sachdev,trugman,kaneleeread,shraimansiggia,%
ruckensteinvarmaschmittrink}
However, even such a basic feature
as the band structure of a single hole has not been experimentally tested.
Indeed, in the low-doping regime the holes are trapped near the
dopant sites.
Even in the translationally invariant case,
the phonon polaron effects~\cite{shuttlezhong,yonemitsubishoplorenzana}
could still destroy the hole dispersion predicted on the
basis of purely electronic models.
Resonant Raman scattering right at half-filling
is sensitive to the details of the dispersion of the fermionic quasiparticles
and thus  provides an experimental information about the single hole
dispersion without having to dope the material.

Before proceeding to the description of our calculations, we would like to make
two comments about the model we use.
First, in this paper we exclusively consider the one-band Hubbard model.
This by itself is a simplification,
since in the ``first-principles" calculation one would start with a
three-band model for the $CuO_2$ unit. Within this model, the lowest gap
(about $2~eV$) is a charge-transfer gap between $Cu$ and $O$
bands.\cite{Varma} However, it is generally accepted
that at energies comparable to this  gap
the hybridization between $Cu$ and $O$ orbitals is relevant, and
one can effectively describe the system by a single degree of freedom per
$CuO_2$
unit, which in turn implies that the three-band Hubbard model can be  reduced
to an effective one-band Hubbard model.\cite{zhangrice,silviaetal,hyber}
The relevant parameter which allows the reduction in the number of degrees
of freedom  is the splitting of the triplet and singlet states of the $O$ and
$Cu$ spins in the Zhang-Rice theory.\cite{zhangrice}
Their estimate of this splitting is several electron volts, and thus
the singlet states relevant to our analysis are well-separated from
the triplet states.
This is corroborated by the measured behavior of the dielectric constant.
For example, in GdCuO$_2$ there are strong absorption features
in the imaginary part of the dielectric constant
around $1.7-2.7~eV$ and around $5~eV$, but only a featureless
continuum in between.\cite{cooperprivate}
We therefore expect that the one-band Hubbard model already captures all
the essential physics of Raman scattering up to photon energies of
order $3-4~eV$.

Our second comment concerns the hopping term in the Hamiltonian.
In this paper, we restrict our model to contain only the
nearest-neighbor hopping, $t$.
Meanwhile, it was suggested~\cite{si} on the basis of the photoemission
data for $YBaCuO_7$, that the next-nearest-neighbor hopping term in the
123 compounds is rather large, $t^{\prime} \sim -0.5t$.
For $La$-based compounds, $t^{\prime}$ was estimated\cite{hyber}
to be $t^{\prime} \sim -0.2t$.
At finite doping, the restriction to only the $t$ term requires care,
as the spin-density wave theory (which we will use in most of our
considerations) is meaningless near the minima of the quasiparticle band,
 unless one takes into account the self-energy
corrections which lift the degeneracy of the quasiparticle spectrum along
the reduced Brillouin zone boundary.\cite{chubukovfrenkel,chub_mus3}
Fortunately, as we will see below, the two-magnon Raman scattering in
cuprates is not a phenomenon that is associated with the minima of
the quasiparticle band, and therefore the details of the quasiparticle
dispersion right near those minima will not be important in our analysis.
For this reason, we believe that for a qualitative analysis of Raman
scattering in cuprates, one can use the simplest model with just the
nearest-neighbor hopping.

The paper is organized as follows.
Sec.~II discusses the general formalism of the two-magnon Raman
scattering in antiferromagnetic insulators.
In Sec.~III, we consider the nonresonant scattering.
In particular, we show how the results of Shastry and Shraiman are
reproduced in the SDW formalism.
We will also discuss the location and shape of the two-magnon peak.
Sec.~IV is the main part of the paper.
We argue that the phenomenon of multiple resonance, described in detail
in that section, is a way to explain the unusual features in
Figs.~\ref{twom} and \ref{peak}.
We study the singularities of the integrals corresponding to that
type of resonant behavior, and present the relevant calculations.
Finally, in Sec.~V, we present a comparison with the experimental
results and  state our conclusions. Some technical details of
our calculations are contained in the Appendix.
Some of the results of this work have been already presented in
a short paper.\cite{prlarticle}

\section{General Considerations}

We start with the general theory of magnetic Raman scattering
in the Hubbard model.
The one-band  Hubbard Hamiltonian is
\begin{equation}
    H= -t\sum_{<i,j>} (c^\dagger_{i,\sigma} c_{j,\sigma} + h.c.)
+ U\sum_i n_{i\uparrow}n_{i\downarrow}.
  \label{add1}
  \end{equation}
A straightforward procedure to derive the coupling of light to the fermions
in this model was recently described by Shastry and
Shraiman.\cite{shastryshraiman1}
In the presence of the slowly varying
vector potential ${\bf A}({\bf x},t)$, each
fermion operator acquires a phase $e^{(i e/\hbar c) \int {\bf A}\cdot
d{\bf l}}$,
and the hopping term gets transformed into
\begin{equation}
H_t = -t\sum_{<i,j>} e^{i \frac{e}{\hbar c} \int^{j}_{i}  {\bf A}\cdot d{\bf
l}}
{}~c^\dagger_{i,\sigma} c_{j,\sigma}  + h.c.
\label{n1}
\end{equation}
The vector potential is supposed to vary slowly on lattice scales, and one
can then approximate the phase factor by
\begin{equation}
\int^{j}_{i}  {\bf A} ( {\bf l} ) \cdot d {\bf l} \approx
{\bf A}\left(\frac{i+j}{2}\right)\cdot {\bf R}_{ij}.
\label{n2}
\end{equation}
Substituting this into the expression for $H_{t}$ and expanding in powers
of the vector potential, one obtains upon transforming to the momentum space
\begin{equation}
H_t =  H^{ {\bf A}={\bf 0} }_{t} -  \frac{e}{\hbar c} \sum_q {\bf j}_{q}
\cdot {\bf A}_{-q} + \frac{1}{2} ( \frac{e}{\hbar c})^2
\sum_{q_1} \sum_{q_2}
{\bf A}_{-q_1}\cdot
{\bf \tau}_{q_1+q_2}
\cdot {\bf A}_{-q_2},
\label{n3}
\end{equation}
where the current and the stress-tensor operators are
\begin{eqnarray}
j^{\alpha}_{q} &=& \sum_{k} \frac{\partial \epsilon_k}{\partial k_{\alpha}}
c^{\dagger}_{k + q/2, \sigma}~c_{k -q/2, \sigma}, \nonumber \\
\tau^{\alpha,\beta}_{q} &=& \sum_{k} \frac{\partial^2 \epsilon_k}{\partial
k_{\alpha} \partial k_{\beta}} c^{\dagger}_{k
+ q/2, \sigma}~c_{k -q/2, \sigma},
\label{n4}
\end{eqnarray}
and $\epsilon_k$ is the electron dispersion,
  \begin{equation}
\epsilon_k=-2t(\cos k_x +\cos k_y).
\label{n44}
\end{equation}
The vector potential can be quantized in the  usual way
\begin{equation}
{\bf A}_{q,\lambda}= g_q
({\bf \hat e}_\lambda l_{-q}+{\bf \hat e}^{\ast}_\lambda l^\dagger_q),
\label{n5}
\end{equation}
where $g_q = (2 \pi \hbar c^2 /(\omega_q V))^{1/2}$ and $\omega_q$ is
the light frequency.
The velocity of light is several orders of magnitude larger than the
Fermi velocity, and therefore we can safely set the momenta of
photons equal to zero.

Consider now separately the last two terms in Eq.~(\ref{n3}), which
contain the vector potential.
The last term  is already quadratic in $\bf A$, so
for photon scattering we have to consider only the first-order
contribution from it.
This contribution does not lead to resonant scattering and,
at least in the resonant regime, can be neglected compared
to the resonant scattering from the first term.
For this reason we neglect the last term in Eq.~(\ref{n3}) in this paper,
and focus only on the first term which leads to photon scattering in the
second order of perturbation theory. This second-order scattering process
involves an intermediate particle-hole state of the fermionic system which
can emit or absorb collective bosonic excitations before collapsing
into the outgoing photon.
In particular, in the two-magnon resonant scattering, a photon with the
energy $\omega_i$ is injected into the antiferromagnetically ordered set
of electrons.
This photon creates a virtual particle-hole pair which then emits two
spin-waves with momenta $\bf k$ and $\bf -k$, and then annihilates into
an outgoing photon with the energy $\omega_f$.

Ignoring the polaritonic effects,
the Raman scattering cross section\cite{hayesloudon}
is obtained from the lowest-order Golden Rule
\begin{equation}
R = \frac{8\pi^3 e^4}{\hbar^3 V^2 \omega_i \omega_f}
{}~\sum_{i,f} |<f|M_R|i>|^2 ~\delta(\hbar \omega_{i} - \hbar \omega_{f} +
\epsilon_i - \epsilon_f),
\label{intensity}
\end{equation}
where $\epsilon_i$ and $\epsilon_f$ are the energies of the initial and final
states of the system ($\epsilon_f - \epsilon_i =\Omega$
is a total frequency of two magnons in the final state), and the summation
over the final states $f$ includes the integration over the photon momenta.
Further, $<f|M_R|i>$ is a matrix element, given by
\begin{equation}
M_R \equiv <f|M_R|i> = \sum_n\Big[
{<f|{\bf j}_{k_f}\cdot{\bf\hat e}^{\ast}_f|n>~
<n|{\bf j}_{-k_i}\cdot{\bf\hat e}_i|i>
\over
\epsilon_i+\hbar \omega_i-\epsilon_n+i\delta
}+
{<f|{\bf j}_{-k_i}\cdot{\bf\hat e}_i|n>~
<n|{\bf j}_{k_f}\cdot{\bf\hat e}^{\ast}_f|i>
\over
\epsilon_i-\epsilon_n-\hbar \omega_f+i\delta
}
\Big].
\label{matrix}
\end{equation}
Here the summation is over the intermediate electronic states, labeled as $n$.
Our primary goal will be to calculate the dependence of this matrix
element on the incident photon frequency.

For completeness, we also list a number of possible experimental scattering
geometries.\cite{shastryshraiman1}
They differ in the polarizations of the incident (${\bf\hat e}_i$) and
scattered (${\bf\hat e}_f$) photons.
For the linearly polarized light, the scattering geometries are
\begin{eqnarray}
A_{1g} &:& {\bf\hat e}_i = \frac{{\bf \hat x} +{\bf \hat y}}{\sqrt{2}},
{}~~{\bf\hat e}_f = \frac{{\bf \hat x}+{\bf \hat y}}{\sqrt{2}}; \nonumber
\\
B_{1g} &:& {\bf\hat e}_i = \frac{{\bf \hat x} +{\bf \hat y}}{\sqrt{2}},
{}~~{\bf\hat e}_f = \frac{{\bf \hat x}-{\bf \hat y}}{\sqrt{2}}; \nonumber
\\
B_{2g} &:& {\bf\hat e}_i = {\bf \hat x}, ~~ {\bf\hat e}_f = {\bf \hat y}.
\label{geom}
\end{eqnarray}
For the circularly polarized light, the scattering geometries are
\begin{eqnarray}
LL &:& {\bf\hat e}_i = \frac{{\bf\hat x} +i {\bf\hat y}}{\sqrt{2}},
{}~~{\bf\hat e}_f = \frac{{\bf\hat x} +i {\bf\hat y}}{\sqrt{2}}; \nonumber
\\
LR &:& {\bf\hat e}_i = \frac{{\bf\hat x} +i {\bf\hat y}}{\sqrt{2}},
{}~~{\bf\hat e}_f = \frac{{\bf\hat x} -i {\bf\hat y}}{\sqrt{2}}.
\label{circgeom}
\end{eqnarray}
The scattering in each geometry measures a particular combination of the
components of the scattering tensor. It is convenient to
decompose $<f|M_R|i>$
into four one-dimensional representations of the 2D square lattice
symmetry group $D_{4h}$:
\begin{eqnarray}
M^{A_1}_{R} &=& \langle x|M_R|x\rangle + \langle y|M_R|y\rangle, \nonumber \\
M^{A_2}_{R} &=& \langle x|M_R|y\rangle - \langle y|M_R|x\rangle, \nonumber \\
M^{B_1}_{R} &=& \langle x|M_R|x\rangle - \langle y|M_R|y\rangle, \nonumber \\
M^{B_2}_{R} &=& \langle x|M_R|y\rangle + \langle y|M_R|x\rangle.
\end{eqnarray}
Then, the scattering in the experimental
$A_{1g}$ geometry measures a combination of $M^{A_1}_{R}$
and $M^{B_2}_{R}$, the scattering in the $B_{1g}$ geometry measures
a combination of $M^{B_1}_{R}$ and $M^{A_2}_{R}$, etc.
In the two-magnon Raman scattering, the energy transfer
$\omega_i - \omega_f$ is small compared to the gap between conduction and
valence bands.
In this situation, the dominant components of the scattering tensor are
found experimentally to be those of $M^{A_1}_{R}$ and $M^{B_1}_{R}$
symmetry.\cite{askandrey2}
It is therefore sufficient to study scattering in the $A_{1g}$ and
$B_{1g}$ experimental geometries.

The states in the general expression of Eq.~(\ref{matrix}) for the matrix
element are the full many-body states of the system, which contain both
the electronic and spin excitations.
Therefore, to fully describe the two-magnon Raman scattering we will need
two different types of vertices.
The first is the interaction between the vector potential of light and
the fermionic current density -- this vertex is given in Eq.~(\ref{n3}).
The second is the interaction between the fermions and the spin-waves.

A convenient way to express both types of vertices on equal footing
is to use the spin density wave (SDW) formalism\cite{schriefferwenzhang}
to describe the electronic state at half-filling and the excitations
around it. In the SDW formalism, one introduces a long-range order in
${\bf S}_q = \sum_k ~c^{\dagger}_{k +q, \alpha} \sigma_{\alpha,\beta}
c_{k,\beta}$ with ${\bf q = Q} \equiv (\pi,\pi)$ and uses it to decouple
the Hubbard interaction term.
The diagonalization then yields two bands of electronic states (the conduction
and valence bands) with the gap $2\Delta \sim U$ in the strong coupling
limit that will be assumed throughout this work.

In terms of the conduction and valence band quasiparticle operators
$a^\dagger_{k\sigma}$ and $b^\dagger_{k\sigma}$,
the quadratic part of the Hubbard Hamiltonian takes the form
  \begin{equation}
H={\sum_{k\sigma}}' E_k (a^\dagger_{k\sigma} a_{k\sigma}-
b^\dagger_{k\sigma} b_{k\sigma}),
  \label{add2}
  \end{equation}
where the prime restricts summation to the magnetic Brillouine zone,
and the quasiparticle energy is $E_k=\sqrt{\epsilon_k^2 + \Delta^2}$.
We will also need the relation between the new and the old
quasiparticle operators.
It reads
\begin{eqnarray}
c_{k,\sigma} &=& u_k a_{k,\sigma} + v_k b_{k,\sigma}, \nonumber \\
c_{k +Q,\sigma} &=& sgn(\sigma) (u_k b_{k,\sigma} - v_k  a_{k,\sigma}),
\label{bogol}
\end{eqnarray}
where the fermionic Bogolyubov coefficients are
  \begin{equation}
u_k =\sqrt{{1\over 2}\left(1+{\epsilon_k\over E_k}\right)},~~
v_k =\sqrt{{1\over 2}\left(1-{\epsilon_k\over E_k}\right)}.
  \label{uv}
  \end{equation}
Note that in the large-U limit, $2u_k v_k\simeq 1$.

We now discuss the vertices. The current operator can be rewritten as
  \begin{equation}
{\bf j}_{q=0} =
\sum_k{\partial \epsilon_k\over\partial {\bf k} }
c^\dagger_{k\sigma} c_{k\sigma}=
{\sum_k}'{\partial \epsilon_k\over\partial {\bf k} }
(c^\dagger_{k\sigma} c_{k\sigma}-
c^\dagger_{k+Q,\sigma} c_{k+Q,\sigma}).
  \label{j}
  \end{equation}
Upon performing the Bogolyubov transformation, we obtain the current
in terms of valence and conduction band fermions,
  \begin{eqnarray}
{\bf j}_{q=0} =
{\sum_{k\sigma}}'  {\partial \epsilon_k\over\partial {\bf k} }
\Big[
  &(& 2u_k v_k)
(a^\dagger_{k\sigma} b_{k\sigma}+ b^\dagger_{k\sigma} a_{k\sigma})
 \nonumber \\
+ &(& u_k^2-v_k^2)
(a^\dagger_{k\sigma} a_{k\sigma}- b^\dagger_{k\sigma} b_{k\sigma})
\Big].
  \label{fullcur}
  \end{eqnarray}
Note that the terms with only valence or conduction
fermions have a  smallness in $t/U$ from the Bogolyubov coefficients.
Keeping only the leading terms, we obtain
  \begin{equation}
{\bf j}_{q=0} \Rightarrow
{\sum_{k\sigma}}'  {\partial \epsilon_k\over\partial {\bf k} }
(a^\dagger_{k\sigma} b_{k\sigma}+ b^\dagger_{k\sigma} a_{k\sigma}),
  \label{add3}
  \end{equation}
which is the expression we use almost exclusively in this paper.

We can now obtain an expression for the optical conductivity
$\sigma({\bf q}={\bf 0},\omega)$ in the SDW state.
 From Kubo formula, in terms of the particle
current given above,
  \begin{equation}
\sigma(\omega)=e^2{\pi\over\omega}\sum_n
\delta(\epsilon_n-\epsilon_0-\omega)
\left|\langle n |j_x({\bf q}={\bf 0})| 0 \rangle \right|^2,
  \label{conduc1}
  \end{equation}
where $\epsilon_n-\epsilon_0$ is the excitation energy between the ground and
excited SDW states.
At half-filling, we only have to retain the intraband terms in
Eq.~(\ref{fullcur}), and, using also Eq.~(\ref{add2}) for the
valence and conduction fermion energies, we obtain
  \begin{equation}
\sigma(\omega)=e^2{4t^2\pi\over\omega}{\sum_{k\sigma}}'
\sin^2k_x~ {\Delta^2\over E^{2}_k} ~\delta(\omega-2E_k),
  \label{conduc2}
  \end{equation}
which agrees with the result in Ref.~\onlinecite{Bulut}.
We have obtained a compact closed form answer for $\sigma(\omega)$
in terms of complete elliptic integrals.
The answer is nonzero when the photon energy is within the bands,
\begin{equation}
\sigma(\lambda) = {e^2 \over 4\pi\hbar} \left({2\Delta\over\omega}\right)^2
\left[
{4\over\sqrt{\lambda} }
E\left(\sqrt{1-{\lambda\over 4} }\right) -
\sqrt{\lambda} K\left(\sqrt{1-{\lambda\over 4} }\right)
\right], ~~~~0<\lambda<4,
\label{fullcond}
\end{equation}
where $\lambda = (\omega^2 - (2\Delta)^2)/16t^2$ counts the photon
energy from the optical gap edge, and we momentarily restore $\hbar$
for reference purposes.
When $t/U\ll 1$, $2\Delta/\omega \approx 1$, and $\lambda \approx
(\omega -U)U/8t^2$.
The plot of $\sigma(\omega)$ at $t\ll U$ is given below
in Fig.~\ref{newfig}(a).
It has a square root singularity at $\lambda=0^+$
and vanishes as $\lambda\rightarrow 4^-$.
The singularity at $\lambda=0^+$ is due to the degeneracy of the
mean-field dispersion $E_k$ along the reduced Brilliouine zone boundary.
As we described above, this singularity is an artifact of mean-field SDW
approach. When the degeneracy in $E_k$
is broken by either quantum fluctuations or hopping to
further neighbors, only the more usual
logarithmic singularity will survive at $\omega = 2\Delta$. On the other hand,
the vanishing of $\sigma(\omega)$ at $\lambda=4$
is due to the vanishing of the current
form-factor $\sin k_x$ in Eq.~(\ref{conduc2}) at ${\bf k}={\bf 0}$,
which is where the {\it maximum\/} optical gap occurs for SDW states.
There is no mean-field degeneracy at ${\bf k}={\bf 0}$, and thus
$\sigma =0$ at the top of the fermionic band should survive also
beyond the mean-field approximation.

The remaining ingredients for Raman scattering are the magnon propagator
and the magnon-fermion scattering vertex.
In the SDW formalism  the magnons are described as collective modes in
the transverse spin channel.\cite{schriefferwenzhang}
Specifically, the spin-wave excitations correspond to the poles of the
transverse spin susceptibility
\begin{equation}
\chi^{+-}(q,q';t)=i\left\langle T\left(
S^+_q(t) S^-_{-q'}(0)\right)\right\rangle.
\label{add4}
\end{equation}
The total transverse susceptibility is given in the SDW theory by the
diagrams containing a sequence of bubbles made of the conduction
and valence fermions with the four-fermion vertices connecting them.
The restriction to only such diagrams can be justified if one extends
the original $S=1/2$ Hubbard model to large $S$ by considering 2S orbitals
at a given site.\cite{largeSreferences}
At half-filling this extension transforms the Hubbard interaction term into
the Hund's rule coupling which favors the maximum possible spin $S$ at each
site.
Notice that in this situation the mean-field gap between lower and
upper Hubbard bands is related to $U$ as $2\Delta = 2US$. In relating the
large-$S$ calculations to the $S=1/2$ case, it is the gap $\Delta$
which should be kept fixed.

The SDW spin susceptibility has been considered several times in the
literature.\cite{schriefferwenzhang,chubukovfrenkel,askandrey4}
Since the  unit cell is doubled due to the presense of the antiferromagnetic
long range order, we have two susceptibilities --- one with zero transferred
momentum, and one with the momentum transfer ${\bf Q}=(\pi,\pi)$.
The explicit forms of these susceptibilities are
\begin{eqnarray}
\chi^{\pm} (q,q,\omega) &=& - S~
\sqrt{1-\gamma_q\over {1+\gamma_q}}~\left[\frac{1}{\omega - \Omega_q +i\delta}
- \frac{1}{\omega + \Omega_q -i\delta}\right],   \nonumber \\
\chi^{\pm} (q,q+Q,\omega) &=& -S~
{}~\left[\frac{1}{\omega - \Omega_q +i\delta}
+ \frac{1}{\omega + \Omega_q -i\delta}\right].
\label{chi}
\end{eqnarray}
Here $\Omega_q =
4JS\sqrt{1-\gamma_q^2}$ is the magnon frequency, and $J = 4 t^2/(2S)^2 U$ is
the exchange integral.

Further, a sequence of bubble diagrams can be viewed as an effective
interaction between two fermions mediated by the exchange of a spin-wave.
The spin-wave propagators are
$i\langle Te_q(t) e^{\dagger}_q(0)\rangle_\omega =
(\Omega_q - \omega - i\delta)^{-1}$
and
$i\langle Te^{\dagger}_q(t) e_q(0)\rangle_\omega =
(\Omega_q + \omega - i\delta)^{-1}$,
where $e^{\dagger}_{q} (e_q)$ are the boson creation (annihilation)
operators, subindex $\omega$ implies Fourier transform,
and the momentum $\bf q$ runs over the whole Brillouin zone.
A simple experimentation then shows that the forms of the two susceptibilities
are reproduced if one chooses the following Hamiltonian for the interaction
between the original fermionic operators and the magnons:
\begin{equation}
H_{el-mag} = { 2\Delta\over \sqrt{2S} }~\sum_{k}^{\prime} {\sum_q}
{}~\left[\eta_q c^{\dagger}_{k+q,\alpha} c_{k,\beta} (e^{\dagger}_{-q} + e_q) +
{}~{\bar \eta}_q
c^{\dagger}_{k+q,\alpha} c_{k+Q,\beta} (e^{\dagger}_{-q} -
e_q)\right]\delta_{\alpha,-\beta}.
\label{magn-ferm}
\end{equation}
Here and below a prime indicates that the summation is over the reduced
Brillouin zone.
The expressions for $\eta_q$ and $\overline{\eta}_q$ are
\begin{equation}
\eta_q = \frac{1}{\sqrt{2}}\left(\frac{1-\gamma_q}{1+\gamma_q}\right)^{1/4},~~
{\overline\eta}_q = \frac{1}{\sqrt{2}}\left(\frac{1+\gamma_q}{1-\gamma_q}
\right)^{1/4}.
\end{equation}

Performing now the Bogolyubov transformation, we obtain the Hamiltonian
for the interaction between the magnons and the conduction and valence
band fermions \begin{eqnarray}
H_{el-mag} &=&
{}~\sum_{k}^{\prime} {\sum_q}~(a^{\dagger}_{i\alpha
k}a_{i\beta,k+q}e^{\dagger}_{q}
{}~\Phi_{aa} (k,q) +
b^{\dagger}_{i\alpha k} b_{i\beta,k+q}e^{\dagger}_{q}
{}~\Phi_{bb} (k,q) \nonumber \\
&& + a^{\dagger}_{i\alpha k} b_{i\beta,k+q}e^{\dagger}_{q} ~\Phi_{ab} (k,q) +
 b^{\dagger}_{i\alpha k} a_{i\beta,k+q}e^{\dagger}_{q}
{}~\Phi_{ba} (k,q) +~{\rm H.c.}~)~\delta_{\alpha, -\beta}.
\label{tranham}
\end{eqnarray}
To leading order in $t/U$, the vertex functions are given by
\begin{eqnarray}
\Phi_{aa,bb} (k,q) &=& \left[\pm (\epsilon_k+
\epsilon_{k+q})\eta_q + (\epsilon_k-\epsilon_{k+q}){\overline\eta}_q
\right] {1\over \sqrt{2S}},
\nonumber \\
\Phi_{ab,ba} (k,q) &=& 2\Delta~\left[
\eta_q \mp
{\overline\eta}_q \right]{1\over \sqrt{2S}}.
\label{vertices}
\end{eqnarray}

We are now in a position to proceed systematically with the diagrammatic
formulation of the two-magnon Raman scattering.

\section{Nonresonant Scattering}

\subsection{Raman matrix element}
\label{ramanmatrixelement}

We start with the situation when the photon frequencies are much smaller
than the  gap, $\Delta \sim SU$.
In this case, it has been been shown\cite{shastryshraiman1}
that the Loudon-Fleury Hamiltionian\cite{loudonfleury} gives a proper
description of magnetic Raman scattering.
We first show how this Hamiltonian can be reproduced in our
momentum-space diagrammatic formalism.
The key points in the derivation are the following: (i) to leading order
in $t/U$ the current operator necessarily transforms a valence fermion
into the conduction one and {\it vice versa\/};
(ii) the vertex strength for a magnon emission accompanied by the fermion
scattering from the valence to conduction band (and {\it vice versa\/}) is
of order $U$,  while for the fermion scattering {\it within\/} either the
valence or conduction band, the vertex is of order $t$, i.e., much smaller
(see Eq.~(\ref{vertices}));
(iii) for incoming/outgoing photon frequencies smaller than the gap,
all the denominators in the diagrams are of the order of $U$.
We can then identify three simple diagrams contributing towards the matrix
element $M_R$ of Eq.~(\ref{matrix}) to leading order in $t/U$.
They are shown in Fig.~\ref{nonresdiag}.
Notice that while the diagrams of Figs.~\ref{nonresdiag}(a),(b) have
a superficial resemblance to those encountered in the theory of
two-phonon Raman scattering in
semiconductors,\cite{cardonareview} the diagram of Fig.~\ref{nonresdiag}(c)
looks rather different.
We will see shortly that the extra diagram is needed for the answer to
have correct symmetry properties.

Performing the internal frequency integration in these three diagrams, we
obtain
\begin{equation}
M^{(1)}_R  = -4 \mu_q\lambda_q~
{\sum_k}'\left({\partial\epsilon_k\over\partial{\bf k} }
\cdot{\bf\hat e}_i \right)
\left({\partial\epsilon_k\over\partial{\bf k} }
\cdot{\bf\hat e}^{*}_f \right)
\left(
{4\Delta\over 4\Delta^{2}-\Omega^{2}} +
{8\Delta(4\Delta^2 +\Omega^{2})\over (4\Delta^2-\Omega^2)^2}
\right)
\label{dia1}
\end{equation}
for the diagram in Fig.~\ref{nonresdiag}(b),
\begin{equation}
M^{(2)}_R = 4\mu_q\lambda_q{\sum_k}'
\left({\partial\epsilon_k\over\partial{\bf k} }\cdot{\bf \hat e}_i \right)
\left({\partial\epsilon_k\over\partial{\bf k} }\cdot{\bf \hat e}^{*}_f \right)
{8\Delta (4\Delta^2 + \Omega^2)\over (4\Delta^2 -\Omega^2)^2}
  \label{dia3}
  \end{equation}
for the diagram in Fig.~\ref{nonresdiag}(c), and
\begin{equation}
M^{(3)}_R  = 4 \left(\mu^2_q + \lambda^{2}_q\right)~
{\sum_k}'\left({\partial\epsilon_k\over\partial{\bf k} }
\cdot{\bf \hat e}_i \right)
\left({\partial\epsilon_{k-q}\over\partial{\bf k} }
\cdot{\bf \hat e}^{*}_f \right)
 \left( {4\Delta\over 4\Delta^2-\Omega^2}\right)
\label{dia2}
\end{equation}
for the diagram in Fig.~\ref{nonresdiag}(a).
In these three expressions $\Omega$ is a frequency equal to $\omega_i$ or
$\omega_f$.
Specifying it more precisely at this stage is devoid of meaning since
we would then have to take into account subleading diagrams in the
$t/U$ expansion.
We also defined
\begin{equation}
\mu_q = \left[\frac{1}{2} \left(\frac{1}{\sqrt{1 - \gamma^{2}_q}} +
1\right)\right]^{1\over 2};~~\lambda_q = \frac{\gamma_q}{|\gamma_q|}
\left[\frac{1}{2} \left(\frac{1}{\sqrt{1 - \gamma^{2}_q}} -
1\right)\right]^{1\over 2}.
\label{lmu}
\end{equation}
These are related to $\eta_q$ and $\overline\eta_q$ via
\begin{equation}
\sqrt{2}\mu_q=\overline\eta_q+\eta_q;~~\sqrt{2}\lambda_q=
(\overline\eta_q-\eta_q).
\label{lmutoetas}
\end{equation}

Adding up the above expressions, and taking into account that
\begin{eqnarray}
& &{\sum_k}'\left({\partial\epsilon_k\over\partial{\bf k} }
\cdot{\bf\hat e}_i \right)
\left({\partial\epsilon_{k-q}\over\partial{\bf k} }
\cdot{\bf\hat e}^{*}_f \right)\nonumber\\
&=&{\sum_k}'(-2t)^2
\left[ ({\bf\hat x} \sin k_x + {\bf\hat y} \sin k_y)
\cdot {\bf\hat e}_i \right]
\left[ ({\bf\hat x} \sin(k_x-q_x) + {\bf\hat y} \sin(k_y-q_y))
\cdot {\bf\hat e}^{*}_f \right ]\nonumber\\
&=&t^2\left[ e_{ix}e^{*}_{fx}\cos q_x+e_{iy}e^{*}_{fy}\cos q_y\right ],
\label{add5}
\end{eqnarray}
we obtain for the Raman matrix element
\begin{equation}
M_R = - 8t^2 ~\left[ {2 \Delta\over 4\Delta^2 -\omega^2}\right ]
\left[
(e_{ix}e^{*}_{fx}+e_{iy}e^{*}_{fy}) 2\mu_q\lambda_q -
(e_{ix}e^{*}_{fx}\cos q_x +e_{iy}e^{*}_{fy}\cos q_y) (\mu_q^2+\lambda_q^2)
\right].
\label{M_R}
\end{equation}

We now demonstrate that this equation coincides with the Loudon-Fleury
vertex. For this, we first rewrite the Loudon-Fleury Hamiltonian of
Eq.~(\ref{eq1})
in terms of Holstein-Primakoff operators to leading order in $1/S$
\begin{eqnarray}
H_{LF}
&=& \alpha
\sum_{r,\hat\mu=\{\hat x,\hat y\}}
({\bf \hat e}_i\cdot{\bf \hat\mu})
({\bf \hat e}^{*}_f\cdot{\bf \hat\mu})
{\bf S}_r \cdot {\bf S}_{r+\hat\mu}\nonumber\\
&=& \frac{\alpha}{2}
\sum_{r}\left[
e^{*}_{fx} e_{ix}
{\bf S}_r\cdot({\bf S}_{r+\hat x} + {\bf S}_{r-\hat x}) +
e^{*}_{fy} e_{iy}
{\bf S}_r\cdot({\bf S}_{r+\hat y} + {\bf S}_{r-\hat y})
\right ]\nonumber\\
&=& 2S \alpha~
{\sum_q}^{\prime} \Big\{ e^{*}_{fx} e_{ix}
\left[ \cos q_x (a_q a_{-q} + a^\dagger_q a^\dagger_{-q}) +
2 a^\dagger_q a_q \right]\nonumber\\
&+& e^{*}_{fy} e_{iy}
\left[ \cos q_y (a_q a_{-q} + a^\dagger_q a^\dagger_{-q}) +
2 a^\dagger_q a_q \right]
\Big\}.
  \label{H_LF}
  \end{eqnarray}
We then apply the Bogolyubov transformation to the actual magnon operators
that diagonalize the Heisenberg model to first order in $1/S$.
This transformation involves the same factors $\lambda_q$ and $\mu_q$
that we had introduced earlier (in Eq.~(\ref{lmu})) and has the form
\begin{eqnarray}
a^\dagger_q &=& \mu_q e^\dagger_q -\lambda_q e_{-q},\nonumber\\
a_{-q}&=&\mu_q e_{-q}-\lambda_q e^\dagger_q.
\label{uvnew}
\end{eqnarray}
At $T=0$ the only relevant terms in the Loudon-Fleury Hamiltonian
are those containing two magnon creation operators, and restricting
ourselves to those, we obtain the matrix element from the Loudon-Fleury
Hamiltonian in the form
\begin{eqnarray}
M^{L-F}_{R} = 2 \alpha  S
\Big\{&&
e^{*}_{fx}e_{ix}
\left[ \cos q_x(\lambda^2_q + \mu^2_q) - 2\mu_q\lambda_q \right]
\nonumber\\
+&&
e^{*}_{fy}e_{iy}
\left[ \cos q_y(\lambda^2_q + \mu^2_q) - 2\mu_q\lambda_q \right]
\Big\}.
\label{LFbos}
\end{eqnarray}
Comparing the two expressions for the Raman vertex, Eqs.~(\ref{M_R}) and
(\ref{LFbos}), we observe that they coincide if one chooses
$\alpha =  16 t^2 \Delta/[2S (4\Delta^2 -\omega^2)]$. Apart from the
factor $2S$ ($=1$ for $S=1/2$), this is exactly the
expression
which Shastry and Shraiman obtained in their derivation of the
Loudon-Fleury vertex for the Hubbard model.\cite{SScom}

Further, an examination of our diagrammatic derivation
shows that the terms in Eq.~(\ref{M_R}) containing
$\cos q_x$ and $\cos q_y$ correspond to the $S^+S^-$ terms in the
Loudon-Fleury Hamltonian, while the rest of the terms
come from the $S^zS^z$ part.
Going back to Eqs.~(\ref{dia1}), (\ref{dia3}), and (\ref{dia2}) and the
diagrams they correspond to, we can see that the $S^+S^-$ terms
come from the diagram of Fig.~\ref{nonresdiag}(a), which emits
one magnon from the hole line and other from the electron line.
The $S^zS^z$ terms come from the diagrams of
Figs.~\ref{nonresdiag}(b),(c), which emit both magnons either from
the electron or the hole side of the fermionic bubble.
Notice that the ``unusual'' diagram of Fig.~\ref{nonresdiag}(c)
canceled one of the terms in the one of Fig.~\ref{nonresdiag}(b).
Without that cancellation we would not have obtained the correct
Loudon-Fleury expression for the $S^zS^z$ term.
This in particular would have resulted in a violation of the
selection rule prohibiting the $A_{1g}$ scattering.

There is an interesting parallel between this cancellation and
one that occurs in the theory of two-phonon Raman scattering
in semiconductors.\cite{cardonareview} In that case, there is
a large cancellation between the two-phonon terms obtained by the
iteration of a single-phonon---electron vertex, and the diagrams
containing a single two-phonon---electron vertex. The cancellation
in the phonon case is a consequence of the translational symmetry.
In our case, the diagram of Fig.~\ref{nonresdiag}(c) can be seen as
effectively containing a single two-magnon---fermion vertex.
The partial cancellation due to this  diagram  restores
the spin rotation invariance of the Loudon-Fleury Hamiltonian.
It is then reasonable to assume that some cancellations may occur
in higher orders in $t/U$, though we did not check this in explicit
calculations.

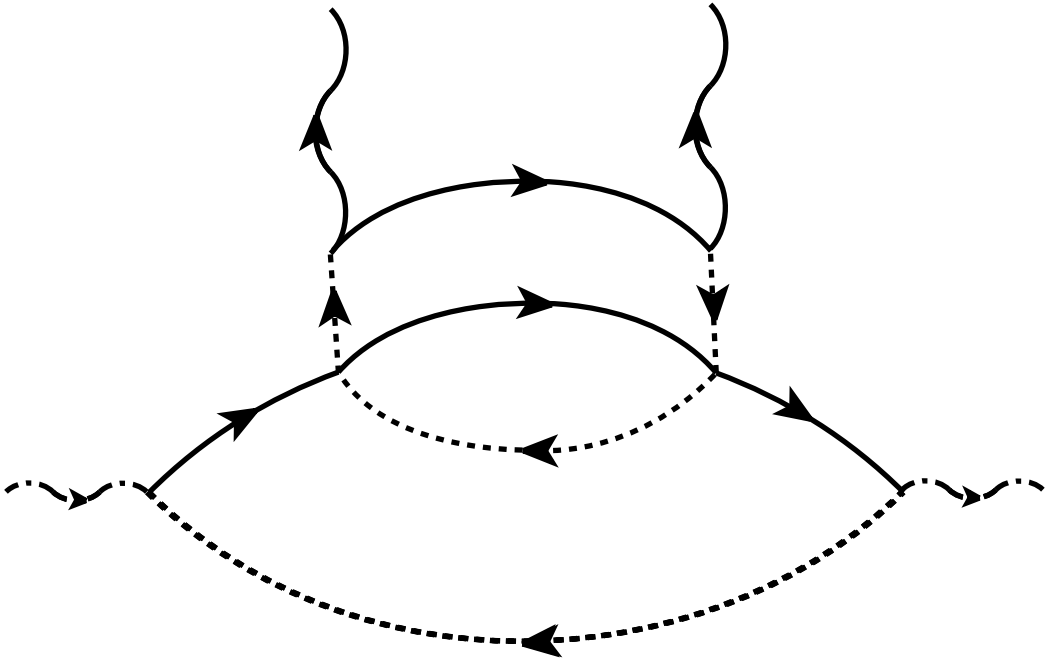
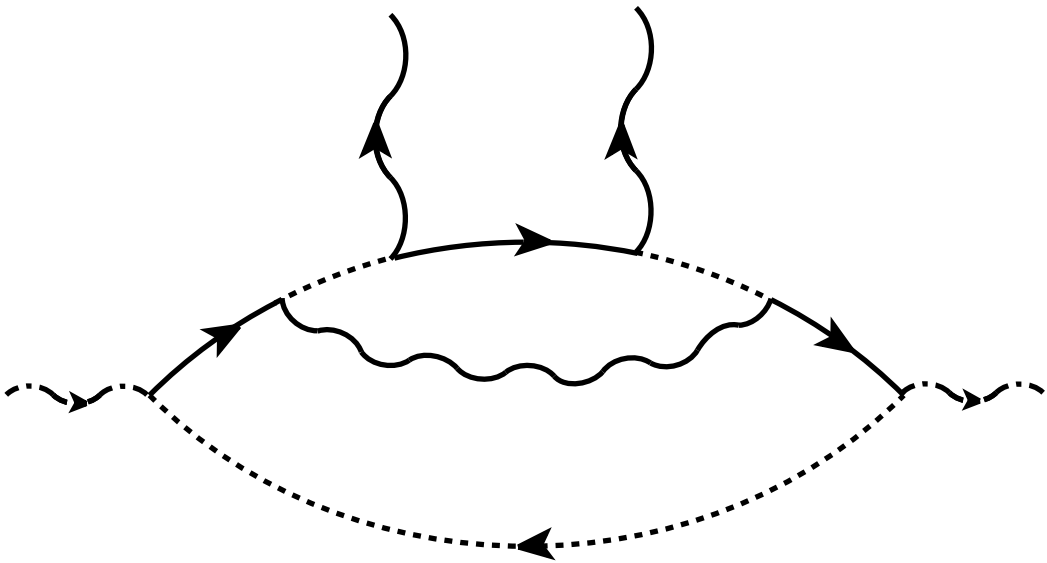
\begin{figure}
\caption{Additional graphs which are of the same order of magnitude
in $t/U$ as those of Fig.~3, but have extra smallness in $1/S$.
The four-fermion vertex is the Hubbard interaction term $U$.
Graph (a) is obtained from Fig.~3(c) by the insertion of a
particle-hole bubble in place of a single four-fermion interaction.
A sequence of such bubbles sums up to an internal magnon line;
an example is given in (b).}
\label{wrongdia}
\end{figure}

Despite the agreement obtained between the two approaches, we are not done yet.
We have so far been performing an expansion in $t/U$, and there exist
other graphs which are of the same order in $t/U$ as the ones retained.
Examples are given in Fig.~\ref{wrongdia}.
However, it is not difficult to observe that all these extra diagrams
have a smallness in $1/S$.
Indeed, each fermion-magnon vertex has an overall factor of $1/\sqrt{2S}$
(see Eq.~(\ref{vertices})).
Associated with each internal magnon line, we have two such vertices,
and {\it no\/} additional summation over the orbitals in the internal
fermionic loop.
Altogether, inclusion of an internal magnon line results in an
extra factor of $1/2S$.
This completes our demonstration of the equivalence of our formalism
to that of Shastry and Shraiman in the non-resonant region.

\subsection{Two-magnon peak}
\label{nonresmagn}

In the preceding subsection, we calculated the Raman matrix element
$M_R$ for the emission of two magnons.
Here we consider the location and shape of the two-magnon peak.
We first observe from Eq.~(\ref{LFbos}) that within the Loudon-Fleury
picture the scattering in the $A_{1g}$ geometry
(${\bf \hat e}_i ={\hat x +\hat y\over \sqrt{2}}$,
${\bf \hat e}_f ={\hat x +\hat y\over \sqrt{2}}$)
vanishes because the Raman vertex $M_R$ is equal to zero.
This result can be expected because at half-filling (and at
large $U$) the Hubbard model is equivalent to the Heisenberg
model, and in the $A_{1g}$ geometry the Loudon-Fleury and
Heisenberg Hamiltonians commute with each other.
On the contrary, in the $B_{1g}$ scattering geometry
(${\bf \hat e}_i ={\hat x +\hat y\over \sqrt{2}}$,
${\bf \hat e}_f ={\hat x -\hat y\over \sqrt{2}}$)
the Raman vertex, as obtained from Eq.~(\ref{LFbos}), is finite
  \begin{equation}
M^{B_1}_{R} \sim (\mu_q^2+\lambda_q^2)(\cos q_x -\cos q_y) =
 {\cos q_x -\cos q_y\over \sqrt{1-\gamma_q^2}}.
\label{B1g}
\end{equation}

Substituting this into the expression for Raman intensity,
Eq.~(\ref{intensity}), we obtain
  \begin{equation}
  R^{B_1}(\omega) \sim
  {\sum_q}' {(\cos q_x -\cos q_y)^2\over 1-\gamma_q^2}
  \delta(\omega-2\Omega_q),
  \label{bare}
  \end{equation}
where $\omega=\omega_i-\omega_f$.
This expression has a divergence near $\omega=8JS$, since the magnon
spectrum is flat and the density of states infinite at the Brillouine
zone boundary.  It is well-known, however, that this result changes
qualitatively when one takes into account the effect of the
magnon-magnon
interactions in the final state.
This was first explained by Elliott {\it et.~al.\/}\cite{elliottetal}
and experimentally verified by Fleury\cite{fleuryexperiment}
for RbMnF$_3$, which has a cubic structure. The calculations were
further extended by Elliott and Thorpe.\cite{elliotthorpe}
They were performed for the two-dimensional case by Parkinson,\cite{parkinson}
and then verified for K$_2$NiF$_4$ by Fleury and
Guggenheim.\cite{fleuryguggenheim}
Since a rather up-to-date treatment of the role of the magnon-magnon
interactions is available in the
literature,\cite{singhreview,canaligirvin}
we merely sketch the relevant calculations with an emphasis on the
$1/S$ expansion, which treats the magnon-magnon interactions in
a systematic way and which, to our knowledge, has not been discussed
before in this context.

The standard way to treat the effects of multiple magnon-magnon
scattering in the Raman problem is to write the Heisenberg
Hamiltonian in terms of the Holstein-Primakoff bosons, diagonalize
it, and keep only the interaction term
$\sim \alpha_k^\dagger\beta_{-k}^\dagger\beta_{-l}\alpha_l$,
which is responsible for multiple scattering of two magnons in the
vacuum (Fig.~\ref{magnon}).
This is the procedure which Parkinson~\cite{parkinson} and others used
to derive the two-magnon Raman intensity.
We note however that for arbitrary $S$, the restriction to only a single
interaction term is not justified, as after the diagonalization of the
quartic term in the Holstein-Primakoff bosons, one also obtains processes
of the form, say,
$\alpha_k^\dagger\beta_{-k}\beta_{-l}^\dagger\alpha_{l}^\dagger$,
which also contribute to the Raman intensity.
The way to avoid this complication is, again, to study the large-$S$
limit.
The key point is that at infinite $S$ the Raman intensity is divergent
at $\omega = 2\Omega_q = 8JS$, which is the maximum possible
two-magnon frequency.
It is achieved when both magnons are right at the Brillouin zone boundary.
At this boundary $\gamma_k =(\cos k_x + \cos k_y)/2 =0$, and the anomalous
term in the bosonic quadratic form in the Heisenberg antiferromagnet vanishes.
In other words, the antiferromagnetic magnons right at the Brillouin zone
boundary behave as free particles.
Now, if $S$ is large, the shift in the two-magnon Raman peak position
due to the magnon-magnon interactions is small, so that one can still
consider magnons as free particles and neglect the anomalous part
in the quadratic form.
In this situation, the only interaction term has two creation and two
annihilation operators.
Moreover, in the Raman problem the total momentum of two magnons is
zero, and we can therefore write the interaction term in the truncated
form
  \begin{equation}
  H_{int}=-{8J\over N}{\sum_k}^{\prime}{\sum_l}^{\prime}
  \gamma_{k-l}\alpha_k^\dagger\beta_{-k}^\dagger\beta_{-l}\alpha_l.
  \label{magn}
  \end{equation}
The prime stands for summation over the magnetic Brillouin zone.
We can further decompose
  \begin{equation}
  \gamma_{k-l}=\gamma_k\gamma_l+
  \tilde{\gamma}_k\tilde{\gamma}_l+
  \overline{\overline{\gamma}}_k\overline{\overline{\gamma}}_l+
  \tilde{\tilde{\gamma}}_k\tilde{\tilde{\gamma}}_l,
  \label{add6}
  \end{equation}
  where the different symmetry factors are
  \begin{eqnarray}
\gamma_k &=& {1\over2}(\cos k_x +\cos k_y),
{}~~\tilde{\gamma}_k = {1\over2}(\cos k_x -\cos k_y), \nonumber \\
\overline{\overline{\gamma}}_k &=& {1\over2}(\sin k_x +\sin k_y),~~
\tilde{\tilde{\gamma}}_k = {1\over2}(\sin k_x -\sin k_y).
  \label{add7}
  \end{eqnarray}
For the $B_{1g}$ scattering geometry, $M_R \propto \tilde{\gamma}_q$, and
hence only $\tilde{\gamma}_k\tilde{\gamma}_l$ survives at
each vertex. The remaining calculation of the series
is straightforward, and we obtain for the Raman intensity
  \begin{equation}
  R_{B_{1g}}(\omega)\sim Im \left[ {I\over 1 + {1\over 4S} I} \right],
\label{ren}
\end{equation}
where
\begin{equation}
I = \frac{1}{N}~{\sum_q}^{\prime} \frac{(\cos q_x - \cos q_y)^2}{\bar{\omega}-
{\bar\Omega}_q+i\delta},
  \label{I}
  \end{equation}
and we introduced the reduced frequencies $\bar{\omega} = \omega/8JS$
and ${\bar\Omega}_q = \Omega_q/4JS$.

The imaginary part of Eq.~(\ref{I}) can be expressed in terms of
complete elliptic integrals,
  \begin{equation}
Im ~I=-{8\bar{\omega}\over\pi\sqrt{1- \bar{\omega}^2}}
\left[ (2-\bar{\omega}^2) K(\bar{\omega}) -2 E(\bar{\omega}) \right].
  \label{Illip1}
  \end{equation}
One can then express $I$ via a dispersion relation,
  \begin{equation}
I(\bar{\omega}) = {8\over\pi^2}\int^1_0 d\nu
{
\nu \left[ (2-\nu^2) K(\nu) -2 E(\nu)
 \right] \over \sqrt{1- \nu^2}~(\bar{\omega}-\nu+i\delta)
}.
  \label{Illip2}
  \end{equation}
This one-dimensional integral is particularly convenient for
a numerical evaluation of $I$.

For $S=\infty$ the denominator in Eq.~(\ref{ren}) is irrelevant, and
this expression reduces to (\ref{bare}), with the only difference
in that we have set $\gamma_q\to 0$ in the $M_R$ of Eq.~(\ref{B1g}),
in order to be consistent with the approximations made in the derivation
of the magnon-magnon interactions.

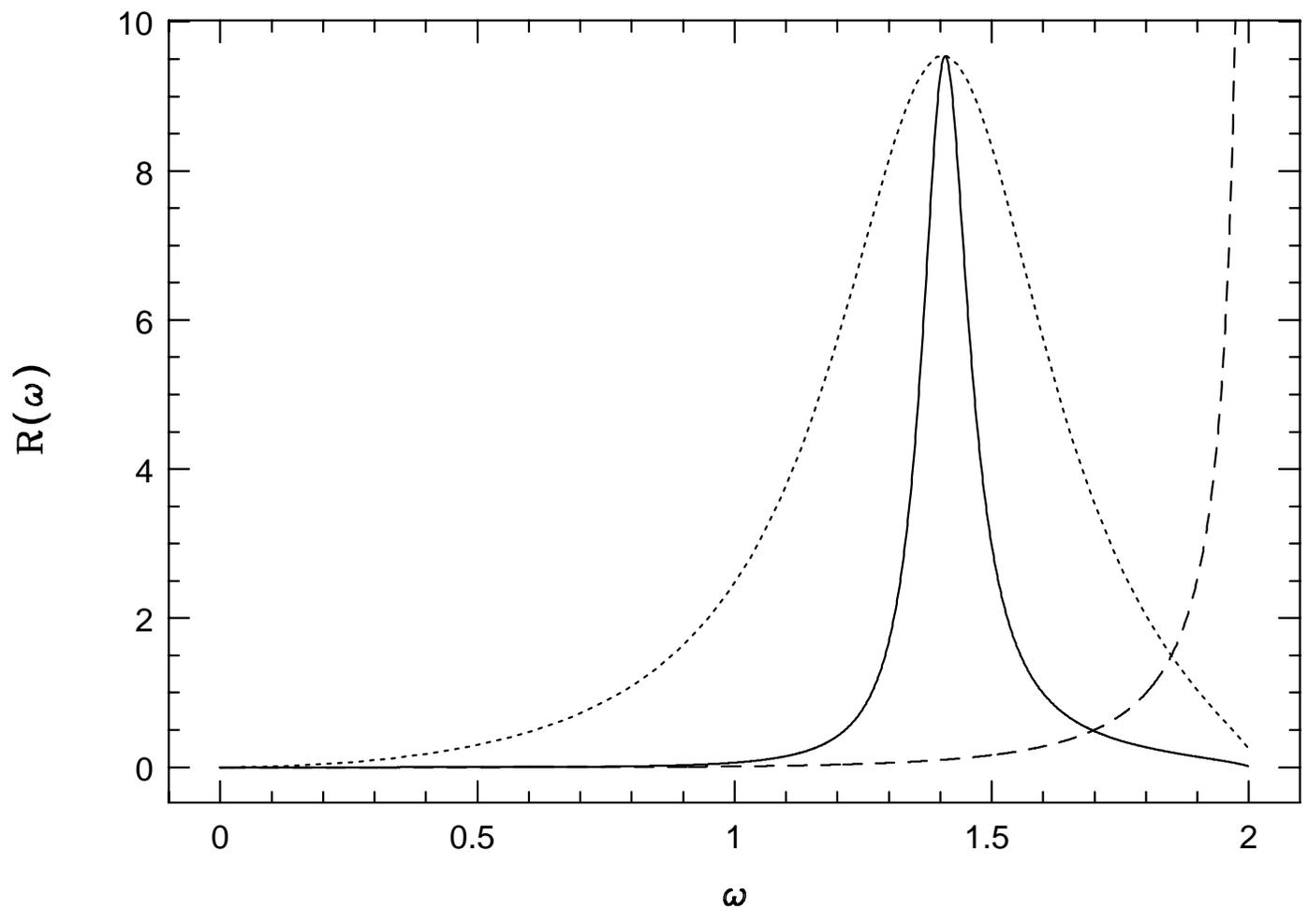
\begin{figure}
\caption{ The profile of the scattering cross-section in the $B_{1g}$
geometry
without the final-state magnon-magnon interactions (dashed line) and with
final-state interaction (solid and dotted line) The solid and dashed lines
were  obtained, correspondingly, by
using the full expression for $I(\omega)$ in (ref{I}) and its expansion
near the top of magnon band which is in line with $1/S$ expansion.
 Notice that while the position of
the two-magnon peak is the same in both cases, the expansion near the top of
magnon band yields much broader two-magnon peak}
\label{magnpeak}
\end{figure}

For large but finite $S$, the peak in the Raman intensity is located close
to but not exactly at $\bar{\omega} =1$.
Expanding  the integral in Eq.~(\ref{I}) in
$1 -\bar{\omega}$, we obtain $I = A +iB$ where
\begin{equation}
A = - 4 \frac{1 - \sqrt{1 - \bar{\omega}^2}}{\sqrt{1 - \bar{\omega}^2}},~~
B = -\frac{4}{\pi}~\frac{1}{\sqrt{1 - \bar{\omega}^2}}~\log(1
-\bar{\omega}^2).
\label{AB}
\end{equation}
Substituting this into Eq.~(\ref{ren}) we obtain after simple manipulations
\begin{equation}
 R_{B_{1g}}(\bar{\omega}) \propto \frac{\sqrt{1 - \bar{\omega}^2}~~\log(1
-\bar{\omega}^2)}{\log^{2}(1
-\bar{\omega}^2) + \pi^2 (1 - (S+1)(1 - \bar{\omega}))^2}.
\label{z}
\end{equation}
We see that when the magnon-magnon interaction is included,
$ R_{B_{1g}}(\bar{\omega})$ no longer diverges but rather has a maximum at
$\sqrt{1 - \bar{\omega}^2} \approx 2\log{(S+1)}/(\pi (S+1))$.
Moreover, right at the zone boundary Raman intensity turns to zero.
If we formally set $S=1/2$ in our $1/S$ expression for Raman intensity,
Eq.~(\ref{z}),
we obtain a peak centered at $\bar{\omega} = 0.696$ (or $\omega
=2.78J$).
The peak position is close to the
$\omega=2.92J$
result obtained by Canali and Girvin.\cite{canaligirvin}
They used the same expression for $I$ as we did and restricted to the
same sequence of bubble diagrams, but they did not make a formal assumption
of large $S$ in their calculations of intensity
and therefore did not expand $I$ in
$1 - \bar{\omega}$ and set $\gamma_q =1$ in (\ref{bare}).
The agreement between our results and theirs shows that the two-magnon peak
position is rather robust. On the other hand, the profile of the two-magnon
peak depends strongly on at which  stage in the calculations
one extends a formal $1/S$ expansion to $S=1/2$. We found that the use of the
$1/S$ expansion results for $A$ and $B$ yield much broader peak than one
obtains using the exact expression for $I$. This last form of the two-magnon
profile is consistent with the results of numerical
calculations~\cite{singhetal} and we  therefore believe that it is closer to
reality.
The profile of the two-magnon intensity is plotted in Fig.~\ref{magnpeak}.

Canali and Girvin also considered the effect of including the
spin-wave velocity renormalization factor into the magnon propagator.
To first order in $1/S$, this renormalization factor effectively
shifts the value of $J$ to $J_{eff} =J (1 + 0.16/2S)$.
The same renormalization must indeed be present in our approach, so the
position of the peak in our theory is at $\omega \sim 3.22J$.

The peculiar features of the two-magnon profile were first reported by
Elliot and Thorpe\cite{elliotthorpe} and Parkinson,\cite{parkinson} who
performed the RPA summation without referring to the $1/S$ expansion,
neglected spin-wave velocity renormalization
and also kept the interaction terms with four creation and annihilation
operators, which were assumed to have the same overall factor
 as in Eq.~(\ref{magn}).
For $S=1/2$, a numerical solution of their equations yields a narrow
peak at $\bar{\omega} = 0.675$ (i.e., $\omega =2.7J$), which is not far from
both our result and that of Canali and Girvin, though we believe that the
restriction to a single interaction term is better justified as long
as the virtual magnons are located near the Brillouin zone boundary.

Experimentally, the two-magnon peak has been observed in $La_2CuO_4$ at
$3000~cm^{-1}$ (see Ref.~\onlinecite{fleury}),
which for $\omega \sim 2.78J_{eff} \approx 3.22J$ yields
 $J \sim0.116~eV$  which is roughly consistent
with $J \sim 0.125~eV$ as inferred from the neutron
scattering\cite{neutron} and NMR data.\cite{NMR}
The Parkinson result corrected by the spin-wave velocity renormalization
($\omega =2.7J_{eff} \approx 3.13J$)
gives somewhat better agreement with the neutron data, but the
difference between the two values of $\omega$ is rather small and practically
irrelevant to subsequent analysis.
In the rest of the paper, we will simply refer to $2.8J_{eff}$ for the
two-magnon peak position.

We now proceed to the central topic of our paper, which is the
discussion of Raman scattering in the resonant regime.

\section{Resonant Scattering}

\subsection{Raman matrix element}

In the section on non-resonant scattering we proceeded systematically
in filtering out the diagrams that were small in $t/U$ or $1/S$.
Our very first step was to observe that the magnon-fermion vertex was
of order $U$ when the fermion scattered between the valence and
conduction bands, and of order $t$ when it scattered within
either the valence or conduction band (see Eq.~(\ref{vertices})).
As the terms in the denominators obtained upon the internal frequency
integration were always of order $U$, all the diagrams with intraband
scattering at any magnon-fermion vertex were small by powers of $t/U$.

It turns out that this property of the denominators is no longer true
in the resonant region, where the photon frequencies $\omega_i$ and
$\omega_f$ differ from the gap $2\Delta$ by quantities of order $JS$.
As a result, some previously omitted diagrams become, as we will see,
not only important, but actually dominant.

Let us first consider the form of the Raman spectrum without the final state
interactions. Then we have
\begin{equation}
R(\omega) \propto \sum_q |M_{R}|^{2} \delta(\omega_i - \omega_f -2\Omega_q)
\end{equation}
Consider now
the graphs with only the intraband scattering at the
fermion-magnon vertices, Fig.~\ref{intraband}.
With all the multiplicity factors included, the internal frequency
integration results in
  \begin{equation}
M^{(4)}_R = -8i{\sum_k}'
{
\left(
{\partial\epsilon_k\over\partial{\bf k} }
\cdot{\bf\hat e}_i
\right)\left(
{\partial\epsilon_{k-q}\over\partial{\bf k} }
\cdot{\bf\hat e}^*_f
\right) [ \mu_q\epsilon_{k-q}-\lambda_q\epsilon_k ]^2
\over (\omega_i-2E_k +i\delta)
(\omega_i-\Omega_q-E_k-E_{k-q} +i\delta)
(\omega_f-2E_{k-q}+i\delta)}
\label{dia5}
  \end{equation}
for the graph of Fig.~\ref{intraband}(a), and
  \begin{equation}
M^{(5)}_R = -8i{\sum_k}'
{
\left(
{\partial\epsilon_k\over\partial{\bf k} }
\cdot{\bf\hat e}_i
\right)
\left(
{\partial\epsilon_k\over\partial{\bf k} }
\cdot{\bf\hat e}^*_f
\right)
[ \mu_q\epsilon_{k-q}-\lambda_q\epsilon_k ]
[ \mu_q\epsilon_k-\lambda_q\epsilon_{k-q} ]
\over
(\omega_i-2E_k +i\delta)
(\omega_i-\Omega_q-E_k-E_{k-q} +i\delta)
(\omega_f-2E_k+i\delta)
}
  \label{dia6}
  \end{equation}
for the graph of Fig.~\ref{intraband}(b).
The relation $\omega_i-2\Omega_q=\omega_f$ is to be remembered here.

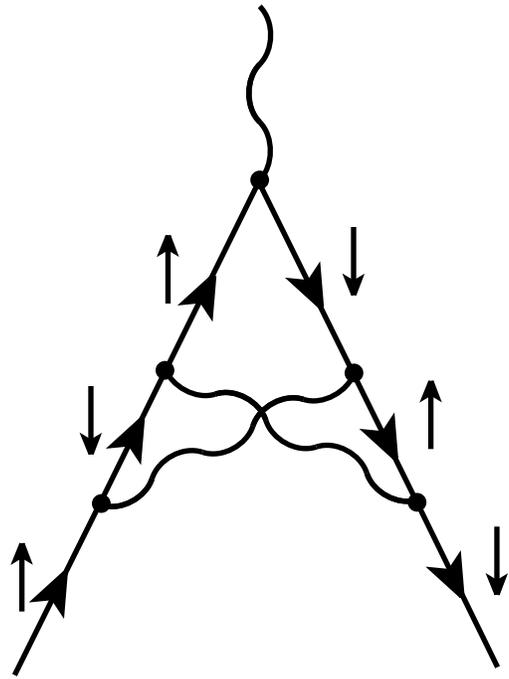
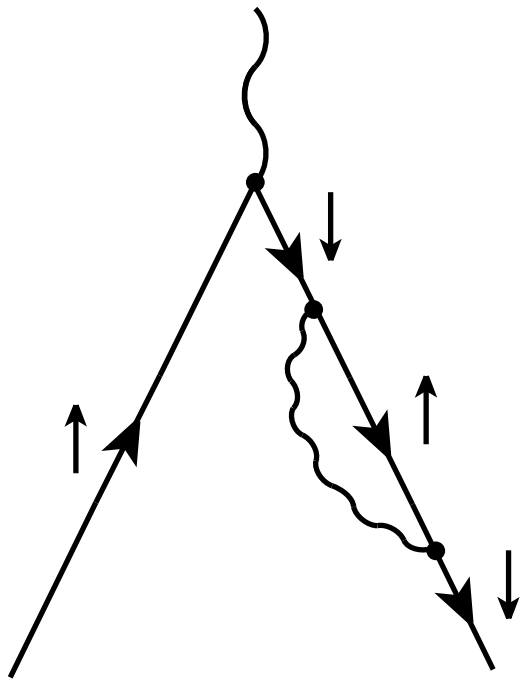
\begin{figure}
\caption{The lowest order vertex (a) and self-energy (b) corrections
to the mean-field theory.}
\label{vertex}
\end{figure}

Without performing the integrals, the naive order of magnitude estimate
is obtained in the resonant region by assuming that the resonant terms in
the denominators are of the order of the bandwidth, i.e., ${\cal O} (JS)$.
Then the two diagrams above are of order $t^4 / (JS)^3$.
The diagrams that were dominant in the nonresonant region
have fewer resonant denominators.
Going back to Eq.~(\ref{M_R}) for the previously considered graphs,
we find that they are of order
$t^2/(2\Delta-\omega) \sim t^2 /JS$.
We see that in the resonant regime, the new diagrams are larger
by a factor $\left(t/JS\right )^2 \sim U/J$.
However, this estimate, while correct for the large-$S$ case, can be
somewhat misleading at the physically relevant value of $S=1/2$, as it
does not take into account strong self-energy and vertex corrections
which are relevant at $\omega \approx 2\Delta$, even though we are
considering the situation exactly at half-filling.
For example, the leading order vertex correction to the magnon-fermion
vertex, shown in Fig.~\ref{vertex}(a), is of order
$t\left(U/JS\right)^2$, whereas the bare vertex is of order $t$.
Simultaneously, the self-energy correction in Fig.~\ref{vertex}(b)
contributes an extra factor of $\left(U/JS\right)$.
Both corrections are small only if we require that $U/JS \ll 1$,
which we indeed do not expect to be satisfied for $S=1/2$.

The form of  the low-energy theory at $U/JS \gg 1$,
has been  discussed by a number of authors. For our considerations, it is
essential that both the numerical~\cite{numericallowenergytheory}
and variational~\cite{analyticallowenergytheory}
studies have found that the resulting quasiparticle Green's function
$G({\bf k},\omega)$ has a quasiparticle pole at low energies,
and that the $k$-space dispersion of that quasiparticle has a width
of order $JS$.
Self-consistent calculations
\cite{kaneleeread,chubukovfrenkel,russianvertexpaper}
also demonstrated that the product of the quasiparticle residue $Z$ and the
effective (renormalized) interaction between quasiparticles $U_{eff}$
scales as $J$ (without corrections this product was equal to $U \times 1 =U$).
In this situation, the relative factor between the new and the old diagrams is
$U_{eff} Z/J = O(1)$. The order of magnitude estimates therefore
 only point to the new diagrams as being
comparable in importance at resonance to those that
fully dominated away from the resonance.

Fortunately, this naive order of magnitude estimate is not the whole story.
When the integrals are actually performed, they may (and as we will see,
they do) yield singular answers for some photon and/or magnon frequencies.
In that case, the most singular integral will dominate.
Notice that because the singularity comes only from the coherent part of the
quasiparticle spectral weight $A({\bf k},\omega)$, we do not need to consider
incoherent part of $A({\bf k},\omega)$, which by itself can be substantial and
spread over the energy scale exceeding $J$ (some studies predict that
$A({\bf k},\omega)$ spreads over the scale of the order
$J\left({t\over J}\right)^\gamma$,
where $\gamma$ is a number $\leq 1$).

We now consider in detail the singular behavior of various diagrams.
The diagrams which contribute to the Loudon-Fleury Hamiltonian have only
singly resonant denominators that lead to singularity in $M_{R}$
 when either the incoming or
the outgoing photon frequency coincides either with the gap, $2\Delta$, or
with the top of the band.\cite{footnoteoncancellationsofsqarepieces}
Near the  gap, the integration over the intermediate momenta
yields $M_{R} \sim |\omega_{i,f} - 2\Delta|^{-1/2}$
for the mean-field form of the quasiparticle dispersion, or an even
weaker logarithmic singularity $M_{R} \sim \log{1/|\omega_{i,f} - 2\Delta|}$
for the renormalized quasiparticle spectrum, which has minima at
$(\pm \pi/2,\pm \pi/2)$ and a parabolic dispersion around the minima.
This last form of the fermionic spectrum was obtained in the numerical,
variational, and perturbative studies of the Hubbard model beyond
the mean-field level.\cite{kaneleeread,chubukovfrenkel,askandrey4}
Near the top of the band (${\bf k} \approx {\bf 0}$), the would be
logarithmic singularity due to the vanishing of the denominator
 is in fact absent because
$\partial \epsilon_k\over\partial {\bf k}$ in the numerator of $M_R$
also vanishes at ${\bf k}={\bf 0}$.

Consider next the diagram in Fig.~\ref{intraband}(a).
Its analytical expression is given by Eq.~(\ref{dia5}).
The denominator in Eq.~(\ref{dia5}) has three terms,
one of which is half of the sum of the other two~\cite{comment}
\begin{equation}
\omega_i - \Omega_q - E_k - E_{k+q} = \frac{1}{2} \left[(\omega_i - 2 E_k) +
(\omega_f - 2 E_{k+q})\right].
\end{equation}
Therefore, if we find $\bf k$ and $\bf q$ such that $\omega_i = 2 E_k$ and
$\omega_f = 2 E_{k+q}$, all three of the denominators in (\ref{dia5}) vanish
simultaneously.
This is known as triple resonance, which is a particular case of the
more general notion of multiple resonance known in Raman
scattering.\cite{cardonareview,martinfalikovreview}
Clearly, a necessary condition for a triple resonance is that the
fermionic bandwidth be larger than the magnon bandwidth.
This condition is satisfied in the mean-field theory, where the
magnon bandwidth is
$4JS$, while the fermionic  bandwidth, found from
  \begin{equation}
E_k \simeq
\Delta+2JS(\cos k_x +\cos k_y)^2,
  \label{add8}
  \end{equation}
is equal to $8JS$. The bandwidths of  ``dressed'' Fermi and Bose
quasiparticles are indeed somewhat different, but we assume that the fermionic
bandwidth is still larger than the magnon bandwidth.

We will present a detailed study of the momentum integration near the
triple resonance in the next subsection, and here merely note that,
unlike its single counterpart,
the triple resonance is not tied to the minimum of either
$E_k$ or $E_{k+q}$.
Thus, unlike for the Loudon-Fleury terms, for triple resonance the
difference between the mean-field and the renormalized forms of the
quasiparticle spectrum does not yield qualitatively different answers
for the singularities.
For this reason, we will keep working with the mean-field form of $E_k$.
In the next subsection we will find that triple resonance does indeed
yield the strongest divergence of the Raman scattering cross-section.

Finally, in the diagram of Fig.~\ref{intraband}(b), we have a product of
$(\omega_i-2E_k)$, $(\omega_f-2E_k)$, and their half-sum in the denominator.
Since $\omega_f=\omega_i-2\Omega_q$, only one of the three terms can
vanish at any time, and this diagram is clearly less singular
than the one with the triple resonance.

In fact, there are also higher order diagrams which have internal
magnon lines or contain four-fermion vertices in low orders.
They may formally be even more singular than the ones above.
However, just as in the nonresonant section, we can omit them
if we assume the large-$S$ limit, which was implicit in the
considerations just outlined.

\subsection{Triple resonance}
\label{triple}

In this subsection, we perform a detailed analysis of the diagram
of Fig.~\ref{intraband}(a), which gives rise to a triple resonance.
The analytical expression for the diagram was given in Eq.~(\ref{dia5}).

It may be useful to start with a  remark on the related calculations of
the two-phonon spectra in semiconductors. In that case the energy scale
of the phonons is quite small compared to the bandwidth in
semiconductors. Therefore an expansion of electronic band structure  to
quadratic order near the minima is {\it a priori\/} warranted, and
the resulting integrals are often doable
analytically.\cite{cardonareview,comm2,martin}

In our case, however, the wavevectors of the magnons contributing
to the two-magnon peak are of the same order as the Brillouin zone
itself, and their energy is comparable to the fermionic bandwidth.
Thus, we are forced to deal with a full spin density wave band
structure without necessarily being able to expand it near some point.
This complicates the integration considerably, although we will
eventually find that some analytical results are still possible.

We now turn to the general analysis of the possible singularities in
Eq.~(\ref{dia5}).

As we mentioned earlier, the three denominators in  Eq.~(\ref{dia5})
vanish simultaneously when the following two conditions hold,
  \begin{equation}
\omega_i = 2E_k;~~~\omega_f = \omega_i - 2 \Omega_q = 2E_{k+q}.
\label{cond}
\end{equation}
Let us call ${\bf k}_0$ the value of $\bf k$ that solves the above
equations, and expand the denominators to linear order about
${\bf k}_0$. To study the conditions for a resonance, we set the numerator
to a constant and obtain
  \begin{equation}
M^{(4)}_R \sim\int d^2k
{1\over
(2{\bf v}_{k_0}\cdot{\bf k} - i\delta)
(({\bf v}_{k_0}+{\bf v}_{k_0-q}) \cdot{\bf k} - i\delta)
(2{\bf v}_{k_0-q}\cdot{\bf k} - i\delta)
}
\label{int}
  \end{equation}
where
${\bf v}_{k_0}={\partial E_k\over\partial {\bf k} }|_{k_0}$ and
${\bf v}_{k_0-q}={\partial E_k\over\partial {\bf k} }|_{k_0-q}$
are the velocities at the two points in the momentum space
where the denominators vanish.

We now immediately see that this integral vanishes unless the two
velocities are strictly antiparallel to each other.
Indeed, let ${\bf v}_{k_0}$ be in the $x$-direction. Then the first term
in the denominator only depends on $k_x$.
If ${\bf v}_{k_0-q}$ has a $y$-component,
then only the second and third terms in the denominator will contain
$k_y$, and both {\it with the same sign\/}. In that case, if we do the
$k_y$ integration first, the integral will vanish since the poles from
both terms will lie in the same half-plane.

The condition for the velocities at ${\bf k}_0$ and
$({\bf k}_0 -{\bf q})$ to be antiparallel has a simple
physical interpretation~\cite{martin}
in the semiclassical picture, where both momenta and coordinates
can be assigned to the quasiparticle wavepackets.
Suppose the incoming photon creates a particle-hole pair at
${\bf k}=\pm {\bf k}_0$.
The energies of particles and holes have different signs, so that the
resulting electron and hole run apart with velocities $+{\bf v}_{k_0}$
and $-{\bf v}_{k_0}$, respectively.
After the emission of two magnons, with the momenta $\pm{\bf q}$,
the electron and hole velocities become $\pm{\bf v}_{k_0-q}$.
Only if the electron and hole are ``aimed back'' at the point of origin,
will they recombine to emit a final photon.
Thus ${\bf v}_{k_0-q}$ has to be antiparallel to ${\bf v}_{k_0}$.

Let us now count the variables in our equations for the resonance.
For fixed $\omega_i$ and $\omega_f$ we have two resonance conditions, an
additional condition that the two velocities be antiparallel, and also the
energy conservation condition $2\Omega_q = \omega \equiv \omega_i - \omega_f$.
Thus, we have four equations for four unknown components of ${\bf k}_0$
and ${\bf q}$.
However, at a fixed $\omega_i$ this system of four equations
does not necessarily have a solution for an arbitrary $\omega_f$.
For example, for the rotationally invariant magnon spectrum and
parabolic fermionic bands, elementary considerations
show that the two conditions in Eq.~(\ref{cond}) plus the
energy conservation condition fully determine
$|{\bf k}_0|$, $|{\bf q}|$, and the angle between them,
so for a fixed $\omega_i$ the condition on velocities
has a solution only at a {\it single\/} $\omega_f$, and
for that one $\omega_f$ {\it all\/} of the magnons are at resonance
at once.
In the general case, the magnon spectrum is not rotationally invariant,
but the triple resonance will still occur only in some range of the
final photon energies $\omega_f$ (and thus only for some magnon energies
$\Omega_q$). We will find that this range is numerically rather narrow in most
cases of interest.

We now study the general solution for triple resonance.
Consider first the  mean-field ($S=\infty$) forms of the fermion and magnon
dispersions. We expand
$2E_k=2\sqrt{\Delta^2+4t^2(\cos k_x + \cos k_y)^2}
\simeq 2\Delta+4JS(\cos k_x + \cos k_y)^2$
and introduce the reduced variables
\begin{equation}
\lambda_i = {\omega_i-2\Delta\over 4JS}, ~~
\lambda_f = {\omega_f-2\Delta\over 4JS},
\label{reduced}
\end{equation}
which count the photon energies from the band edge.
Then Eq.~(\ref{cond}) becomes
  \begin{equation}
\cos k_x +\cos k_y = \sqrt{\lambda_i},~~
\cos l_x +\cos l_y = \sqrt{\lambda_f},
  \label{cond2}
  \end{equation}
where ${\bf k} \equiv {\bf k}_0$, and where we introduced
${\bf l}={\bf k}-{\bf q}$ and also restricted ourselves to
the reduced Brillouin zone.

The condition that the velocities $\partial E_k\over \partial {\bf k}$ and
$\partial E_l\over \partial {\bf l}$ be antiparallel is
\begin{equation}
\sin k_x \sin l_y-\sin l_x \sin k_y =0,
\label{antipa}
\end{equation}
provided that if $\bf k$ is in the first quadrant, then $\bf l$
is in the third (this makes the velocities {\it anti\/}parallel).

Finally, the condition $E_k-E_l=2\Omega_{k-l}$ takes the form
  \begin{equation}
2\sqrt{1-{1\over 4}(\cos(k_x-l_x) +\cos (k_y-l_y) )^2}=\lambda_i-\lambda_f.
\label{b}
  \end{equation}
These four coupled nonlinear equations is the set to deal with.
There exists no general method for solving systems of nonlinear equations.
However, in our particular case, much information can be obtained
analytically by studying two high symmetry directions, which we do next.

Let us first consider the case where $q_x=q_y$.
It is easy to check that for such $\bf q$, the solution for the
resonance has a general form
$k_x=k_y$~($0<k_x<\pi/2$)  and $l_x=l_y$
{}~$(-\pi/2<l_x<0)$. The antiparallel velocities condition
is automatically satisfied, and the remaining conditions yield
  \begin{eqnarray}
\cos k_x &=&{\sqrt{\lambda_i}\over 2},~~~\cos l_x =
{\sqrt{\lambda_f}\over 2},  \nonumber \\
2\sqrt{1-\cos^2(k_x-l_x)}&=&  2\sin(k_x-l_x)=\lambda_i-\lambda_f.
  \label{add9}
  \end{eqnarray}
These three equations trivially lead to
  \begin{equation}
{\bar\Omega}_q = \frac{1}{4}
 \left[\lambda_i -2 + \sqrt{3 \lambda_i (4 - \lambda_i)}\right],
  \label{reso}
  \end{equation}
where as before ${\bar\Omega}_q={\Omega_q\over 4JS}$
is a reduced magnon frequency.

By definition, the upper limit on $\lambda_i$ is equal to 4.
The lower limit is found by locating the point where
$\lambda_f=\lambda_i-2{\bar\Omega}_q$ becomes zero, since
at that point one of the velocities vanishes, and after that we have
parallel rather than antiparallel velocities and thus no contribution
from the triple resonance to the
Raman vertex.
Solving $\lambda_i = 2{\bar\Omega}_q$ together with Eq.~(\ref{reso}) we find
the lower boundary is at ${\bar\Omega}_q=1/2$,  $\lambda_i=1$.
Eq.~(\ref{reso}) is graphically presented in Fig.~\ref{solution}.

Another symmetry direction for which the condition on the velocities is
satisfied automatically is that for which $\bf q$ is along the $x$
or $y$-axis.
In this case we can set $k_y=l_y=q_y=0$, and obtain a set of
equations
  \begin{eqnarray}
\lambda_i&=&(1+\cos k_x)^2,~~~\lambda_f=(1+\cos l_x)^2, \nonumber \\
\lambda_i-\lambda_f&=&
2\sqrt{ 1-{1\over 4}(1+\cos(k_x-l_x))^2 }.
  \label{a}
  \end{eqnarray}
It actually does lead to an analytic answer, but as it is rather
cumbersome, we refrain from presenting it and consider instead
only two special points.
The first one is where  the magnon energy is maximal, i.e.,
$\lambda_i-\lambda_f=2$.
This implies $k_x-l_x =\pi$, and hence  $\lambda_i=9/4$.
The second special point is one at which $\lambda_f=0$, i.e., the
termination point on the lower end. At that point $l_x=-\pi$, and
we easily obtain from Eq.~(\ref{a}) that
  \begin{equation}
  \lambda_i(\lambda_i+1)^2=16,
  \label{add10}
  \end{equation}
with the solution
$\lambda_i\simeq 1.9$.

The whole curve~\cite{extracomm} $\omega_{q} =f(\lambda_i)$ for $q_x =0$
is plotted in Fig.~\ref{solution}.

Another part that is worth deriving analytically is the locus of the
lower and upper termination points of the curves for different
directions of $\bf q$.
It is easy to demonstrate  that for the largest possible
$\lambda_i =4$, the solution for any direction is $\lambda_f =3$,
so that the curves for all directions of $\bf q$ meet at a single
point when $\lambda_i =4$.
To see this, note that $\lambda_i =4$ implies $k_x = k_y =0$.
After substituting this into Eqs.~(\ref{cond2}), (\ref{antipa}),
and (\ref{b}), we obtain the claimed result.

The lower termination point is $\lambda_f=0$, because the triple resonance
clearly requires both the initial and final photon frequencies to lie above
the gap.
This condition corresponds to a straight line given by
$\lambda_i = 2{\bar\omega} = (\omega_i - \omega_f)/4JS$
in the $(\lambda_i, {\bar\omega})$ plane (recall that at resonance,
$\bar\omega = {\bar\Omega}_q$).
This by itself would allow $0<\lambda_i<2$.
A more detailed study shows, however, that the solution only exists
for $\lambda_i>1$.
Notice that curves for different magnon directions terminate at
different points on the line $\lambda_i = 2 {\bar\omega}$, and near
$\lambda_i =1$ only a small fraction of magnon directions, namely
$q_x \approx q_y$, allow a triple resonance there
(see Fig.~\ref{solution}).
At the other end of this boundary (at $\lambda_i =2$), the numerator in
(\ref{dia5}) vanishes.
For all these reasons, Raman intensity is expected to be small in the
region of low $\lambda_i$, and it will not play any role in our
subsequent analysis.

Finally, we can also locate the region where  our set of equations has
a solution for the largest possible magnon frequency $\bar\omega =
{\bar\Omega}_q =1$.
An inspection of Eqs.~(\ref{cond2}), (\ref{antipa}), and (\ref{b})
shows that the solution exists for $2 <\lambda_i <3$.

Combining all these analytical results, we obtain the  allowed region
of triple resonances in the $(\lambda_i,2 {\bar\omega})$ plane.
This region is shaded in Fig.~\ref{solution}.
Notice that although for $2< \lambda_i <3$ the solid line in the figure
(the solution for $q_x =0$) practically coincides with the horizontal
line $ 2 \bar\omega = 2 {\bar\Omega}_q =2$,
 this solid line is actually located {\it below\/}
the maximum two-magnon frequency everywhere except $\lambda_i =9/4$.
As an independent check, we also solved our equations numerically
for a number of directions of $\bf q$ and found that the solutions
were within the shaded region.

The above solution for the triple resonance was obtained for the mean-field
forms of fermionic and magnon dispersions. As we already discussed, quantum
corrections will certainly change the overall scales in the dispersions and
thus modify the resonance conditions. However, we do not expect these
modifications to be substantial especially near the upper termination point
$\lambda_i =4$ as still, at maximum $\lambda_i$ we have $k_x =k_y =0$, and
hence curves with different directions of $q$ will terminate at the same
$\lambda_i$. This region near maximum $\lambda_i$ will play central role in our
subsequent analysis.

Finally, it is worth noticing that the divergence we have found in Raman matrix
element is
an artifact of neglecting the damping of quasiparticles.
If the damping was included, the divergence would be gone, and we would
obtain instead only the enhancement of $M_R$ in the shaded region in
Fig.~\ref{solution}.
For this reason, later in the paper we will refer to an enhancement
rather than to a  singularity in $M_R$.
It is important, however, that up to some damping levels the enhancement
of $M_R$ is going to be substantial, and the singularity analysis is a
useful guide to the actual physics.

\subsection{Two-magnon scattering}

So far we have considered the behavior of the Raman matrix element
as a function of incoming frequency, neglecting finite state interaction,
 and selected the diagram which
gives rise to a strong enhancement of $M_R$ in some range of outgoing photon
frequencies for a given $\omega_i$.
 At the same time,
 as we discussed in Sec.~\ref{nonresmagn},
the density of states is divergent at the magnetic zone boundary,
and in the absence of finite state interaction
this naturally gives rise to a divergence of $R$ at $\omega = 8JS$.
For $S=1/2$, which we will assume below, this divergence is at  $\omega =4J$.

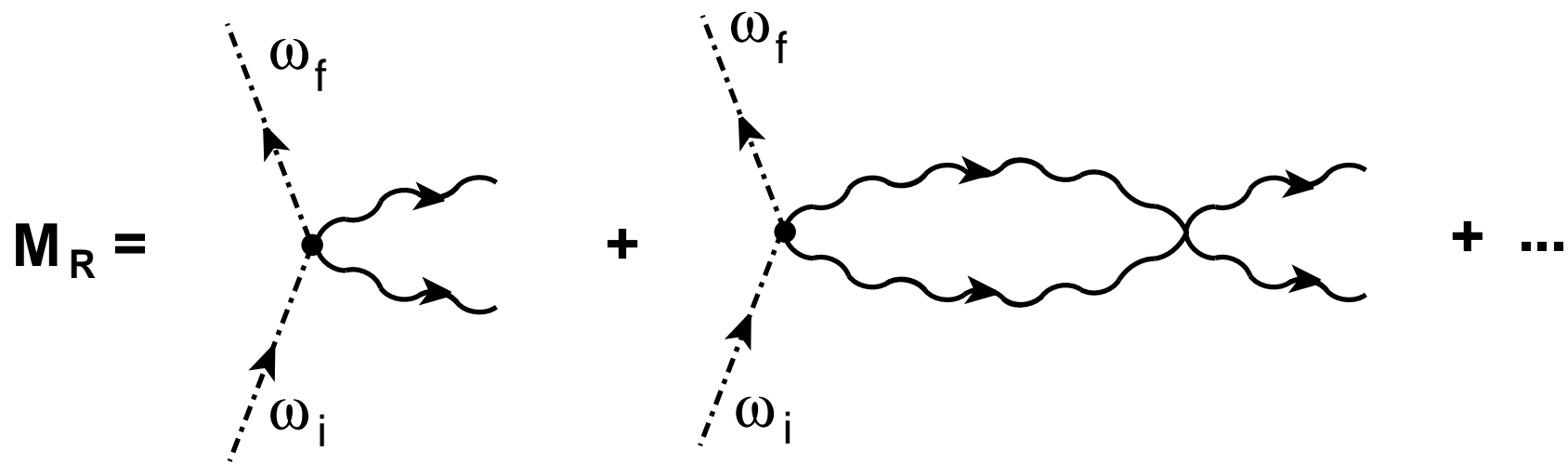
\begin{figure}
\caption{A series of diagrams with the
final-state magnon-magnon interactions.
The black dot is the Loudon-Fleury vertex for the photon-magnon
interaction.}
\label{magnon}
\end{figure}

If we now switch on the magnon-magnon interactions, the divergence at $4J$
will be gone, and we will obtain instead the maximum of $R$ at a smaller
frequency.
In principle, this frequency should be different from that obtained in
Sec.~\ref{nonresmagn}, as the Raman matrix element in Eq.~(\ref{dia5})
has a complicated dependence on the magnon momentum $\bf q$, and the
terms in the resulting series problem cannot be factorized.
On the other hand, the location of the two-magnon peak at about $3J$
is in agreement with the experimental\cite{fleury,askandrey2}
as well as numerical
data.\cite{singhreview,singhetal,norigalianobacci}
If so, the influence of the $\bf q$-dependence of $M_R$ on the
magnon-magnon scattering should not be substantial.
There will be also a feedback effect from the magnon-magnon scattering
on the location of the region where the Raman matrix element is
enhanced due to triple resonance.
However, this feedback effect will be small in the semiclassical (large $S$)
approximation, which, as we discussed in Sec.~\ref{nonresmagn}, is likely to
work well even for $S=1/2$.  For all these reasons, for the rest of our
discussion we adopt a semiphenomenological approach and assume that at
any given $\omega_i$ the Raman spectrum $R(\omega)$ has two
{\it independent\/} peaks: one is due to the triple resonance, which is
particularly relevant to the first diagram in Fig.~\ref{magnon}
which has no finite state
interaction, and
the other, which for definiteness
we assume to be at $\omega =2.8J_{eff}$, is due to the magnon-magnon
scattering (all other diagrams in Fig.~\ref{magnon}).
 In essence, the approximation we made implies that we are considering an
``effective'' Loudon-Fleury-like model, in which the Raman vertex is
considered as constant in the calculations of the two-magnon peak position,
but  the overall magnitude of the vertex does indeed strongly
depend on how close we are to the triple resonance region.

Without considering in detail the effects of damping, we cannot
conclude which of the two peaks is stronger.
The experiments seem to indicate that the peak at $2.8J_{eff}$ is stronger,
and the enhancement of the Raman matrix element can only be responsible
for the asymmetric ``shoulder-like" behavior of the two-magnon profile
(see the discussion in Sec.~\ref{dis}).
At the same time, if  we fix $\omega$ at $2.8J_{eff}$,
 as in Fig.~\ref{peak}, and
consider the variation of the two-magnon peak intensity as a function
of the incident photon frequency $\omega_i$ (or $\lambda_i$), this
intensity will clearly have a maximum where an $\omega=const$ line
intersects the region of the triple resonance.

To see where the maximum occurs, we neglect for simplicity the
renormalization of $J$ due to quantum fluctuations and draw a
horizontal line in
Fig.~\ref{solution} at the reduced frequency $\bar{\omega} = 0.7$
(we checked that the results below are practically insensitive to whether
we use $J$ or $J_{eff}$ in the magnon spectrum; however, if we used $J_{eff}$,
we would also have to consider the $1/S$ renormalization for the fermionic
spectrum, which does not lead to new physics but substantially complicates the
calculations).
We see from the figure that the line at $\bar{\omega} = 0.7$
intersects the region of triple resonances in two places.
One occurs at $\lambda_i$ very close to 4, i.e., when the incident photon
is near the {\it top\/} of the fermionic band.
The other occurs for $\lambda_i$ close to 1, i.e., not far from the
{\it bottom\/} of the fermionic band.
We know from the discussion given above that the region $1<\lambda_i<2$,
especially for $\lambda_i$ close to 1, is rather ``esoteric" in that only
a small fraction of magnon directions
 allow for a triple resonance there (see Fig.~\ref{solution}).
On the other hand, for $\lambda_i$ close to 4, we will encounter no
such problem, and we thus expect a much larger enhancement there.
Thus, we focus entirely on the latter value of $\lambda_i \approx 4$.

We now consider how the two-magnon peak amplitude increases as $\omega_i$
approaches the critical value where the triple resonance and the
magnon-magnon peak occur at the same value of $\bar{\omega}$.
First, since we are close to the top of the fermionic band, we can
perform a quadratic expansion of the band structure near the maximum.
Second, we see from Fig.~\ref{solution} that near $\lambda_i =4$, the
triple resonance region is  very narrow, so that to first approximation
we can consider that the triple resonance occurs only along a single line
in the $(\lambda_i,\bar{\omega})$ plane.
As we discussed in Sec.~\ref{triple}, this would have been the case if
the magnon spectrum was rotationally invariant, and one can easily
check that it approximately is for ${\bar\Omega}_q \sim 0.7$.
Accordingly, we expect it to be a good approximation if we
linearize the magnon spectrum around ${\bf q}={\bf 0}$,
  \begin{equation}
  {\bar\Omega}_q\simeq {q\over\sqrt{2}}.
  \label{add11}
  \end{equation}
The integral for the triple resonance can now be done analytically.
The details of the calculations are presented in the Appendix.
Here we quote the results.
The resonance is allowed at the line
\begin{equation}
\lambda_i=4-(\bar{\omega}-{1\over2})^2,
\label{lres}
\end{equation}
which, in accordance with Fig.~\ref{solution}, terminates at
$\lambda_i =4$ for $\bar{\omega} =1/2$, and intersects the two-magnon peak
position ($\bar{\omega }= 0.7$) at $\lambda_i = \lambda^{res}_i =3.96$.
The maximum of the two-magnon peak amplitude thus corresponds to the
incoming photon frequency $\omega_i$ that puts the particle-hole pair
members very close to the tops of their respective bands.

Let us now fix the two-magnon frequency at the two-magnon peak position
and vary $\lambda_i$ (as in Fig.~\ref{peak}).From
the results in the Appendix, the Raman matrix element is then
  \begin{equation}
M_R \sim {4 - \lambda_i \over (\lambda_i-\lambda^{res}_i+i\delta)^{3/2}
}
\label{si}
  \end{equation}
The factor in the numerator comes from the vanishing of the term
$\partial \epsilon_k\over\partial {\bf k}$ in the numerator of
Eq.~(\ref{dia5}) at ${\bf k}={\bf 0}$, which is the value of $\bf k$ of the
resonant particle-hole pair created by the incoming photon
for $\lambda_i=4$, and also due to the fact that $\lambda_i =4$
corresponds to ${\bar\Omega}_q-{1\over 2}=0$ in which case  the overall factor
$(1-{{\bar\Omega}_q\over q^2}) \equiv (1 - {1\over {2 {\bar\Omega}_q} })$
 in Eq.~(\ref{A3}) in the Appendix vanishes.

For $\lambda_i$ very  close to $\lambda_i^{res}$, the term in the numerator
 can be considered as a constant, and we obtain
\begin{equation}
M_R \sim {1\over [\lambda_i-\lambda_i^{res}+i\delta]^{3/2} }.
  \label{sing}
  \end{equation}
In practice, however, the damping of fermions prevents the true singularity at
$\lambda_i = \lambda_i^{res}$, and the increase of $M_R$ can be measured at
some distance away from $\lambda_i^{res}$. In this situation, the difference
between $\lambda_i^{res}$ and $4$ can be neglected, and we obtain from
(\ref{si})
  \begin{equation}
  M_R\sim {1 \over \left(\lambda_i-\lambda_i^{res}+i\delta\right)^{1/2}},
  \label{redsing3}
  \end{equation}
The sum total of those arguments is that the Raman intensity has
{\it at least \/} an inverse linear singularity from triple resonance.
\begin{equation}
R \sim [\lambda_i-\lambda^{res}_i +i\delta]^{-1}.
\label{res}
\end{equation}
It is important to note that the usual single (Loudon-Fleury) resonance in
2D can {\it at best} give an inverse linear singularity of $R$
for the spin density wave band structure, and only at the mean
field level,
due to the degeneracy of $E_k=\sqrt{\Delta^2+\epsilon_k^2}$.
Once the mean-field degeneracy is broken by a better approximation,
the single resonance is further reduced to a logarithmic singularity.
Therefore the triple resonance singularity is expected to be
at least as strong, and most likely stronger than any at the lower band
edge.

This is to be contrasted with the behavior of the optical conductivity
$\sigma(\omega)$ (see Eq.~(\ref{fullcond}) above and Fig.~\ref{newfig}(a)).
Unlike the Raman cross-section, it  vanishes at the
upper band edge $\lambda_i=4$. Most of the opitical weight
in  $\sigma(\lambda_i)$ is concentrated towards the lower band
edge $\lambda_i=0$, where it has a singularity in the absence of fermionic
damping. Therefore, the optical conductivity resonates
at the lower band edge
while the Raman cross-section is strongest at the upper band edge
of the coherent quasiparticle spectrum.

Notice also that while the Raman intensity is symmetric with respect to the
sign of $\lambda_i - \lambda_i^{res}$, the expression (\ref{si})
for $M_R$, which includes only the leading singularity,
is, strictly speaking, valid only in some region around $\lambda_i^{res}$.
Away from this region, there are no reasons to expect the Raman intensity
to be symmetric. On general grounds, we expect that the intensity
above $\lambda_i^{res}$ must fall down faster than for the same deviation from
$\lambda_i^{res}$ but at $\lambda_i < \lambda_i^{res}$ simply because
no triple resonance is possible for $\lambda_i >4$.

We now discuss in more detail how these (and other) results are related to
experiment.

\section{discussion}
\label{dis}

We first summarize the main results of our work.

We developed a general diagrammatic approach to Raman scattering in
antiferromagnetic insulators.
For photon frequencies small compared to the gap between conduction and valence
band, we
rederived diagrammatically the Loudon-Fleury Hamiltonian,\cite{loudonfleury}
which had been first derived from the Hubbard model by Shastry and
Shraiman\cite{shastryshraiman1} in a different formalism.
We also considered the location and shape of the two-magnon peak by
studying the magnon-magnon scattering in a systematic $1/S$ expansion.
The results of this study are consistent with the earlier considerations by
Canali and Girvin.\cite{canaligirvin}

We studied for the first time the two-magnon Raman scattering in
the so-called resonant regime when the incident and final photon
frequencies are only ${\cal O}(J)$ apart from the Hubbard gap.
In this situation, the diagrams which are subleading in the nonresonant
region become dominant.
We identified the diagram which gave a dominant contribution to the
Raman vertex in the resonant regime, and found the region in
the $(\omega_i, \omega_i - \omega_f)$ plane where the Raman
vertex is divergent, unless one includes into consideration the
damping of quasiparticles.
The divergence is due to the simultaneous vanishing of all three
denominators in the expression for the Raman matrix element.
This phenomenon is called triple resonance.
In the presence of a moderate quasiparticle damping, the divergence
in the Raman vertex is gone, but the enhancement due to the triple
resonance can indeed survive.

Based on the experimental data, we then assumed that the location of the
two-magnon peak in the resonant regime remains practically in the same
place as at small photon frequencies, i.e., at
$\omega_i - \omega_f \sim 3J$.
For nearly all incoming photon frequencies $\omega_i$ falling within
the Hubbard bands (i.e., corresponding to the features in the optical
absorption spectrum), this magnon frequency is located below the region
where our calculations show a triple resonance is allowed
(see Fig.~(\ref{solution})).
Based on these facts, we formulated an ``effective'' Loudon-Fleury-like
theory in which the position and shape of the Raman profile in the
two-magnon peak area of $\approx 3000~cm^{}$ (but not at
higher magnon frequencies) is solely due to the final state
magnon-magnon interactions, while the triple resonance in the
Raman vertex at larger $\omega \sim  4000~cm^{-1}$
leads to the ``shoulder-like'' behavior in the magnon profile.
When the incident photon frequency approaches $\omega^{max}_i = 2\Delta + 8J$,
which corresponds to the particle and the hole being excited to nearly the
tops of their respective
bands, the triple resonance region allowed at a given
$\omega_i$ shifts down to lower transferred
frequencies, and finally intersects
the value of $\omega$ where the two-magnon peak
is located. We found that this happens when $\omega_i /2$
is very close to the
top of the quasiparticle band,
$\omega_i = \omega^{res}_i \approx \omega^{max}_i$. Right at the
crossing point, the  intensity of the
two-magnon peak amplitude
diverges (in the absence of quasiparticle damping).

As $\omega_i$ approaches this critical value from below, the Raman
intensity increases as $R \sim (\omega^{res}_i - \omega_i)^{-1}$ at some
distance away from the resonance, and as
$R \sim (\omega^{res}_i - \omega_i)^{-3}$ in the immediate vicinity of the
resonance.
In a real situation, the true divergence  will be gone due to damping
of quasiparticles, and the two-magnon peak intensity as a function
of $\omega_i$ will instead have a maximum at $\omega^{res}_i$.
We therefore do not expect to see cubic dependence, but inverse
linear increase of $R$ should be observable.

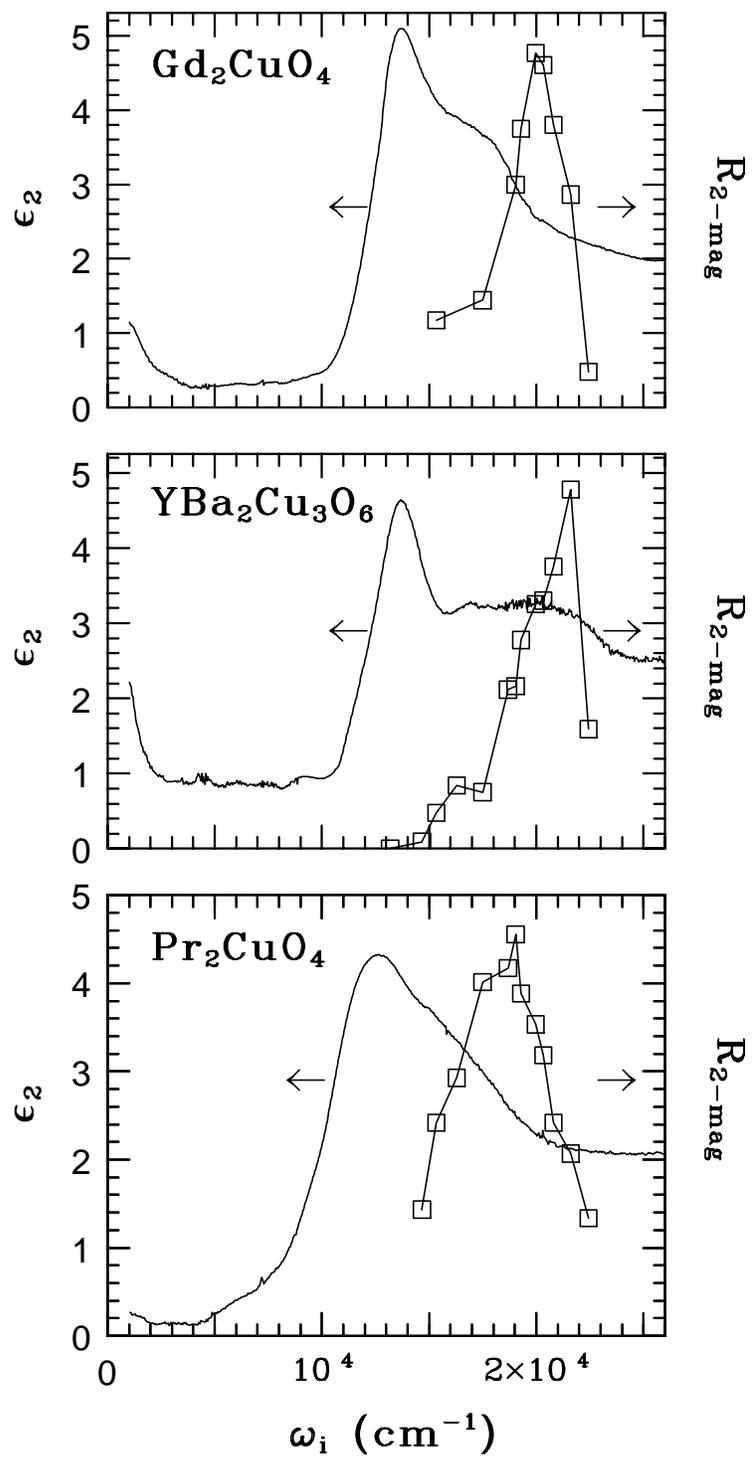
\begin{figure}
\caption{The strength of the two-magnon peak as a function
of incoming photon frequency.
Also shown is the imaginary part of the dielectric constant.
Data courtesy of the authors of Ref.~\protect\onlinecite{ranliu}.
The data in the Figure were obtained at room temperatures. At low $T$,
both the peak in the optical absorption and the Raman profile
shift to higher frequencies by about
$0.1 eV$.\protect\cite{kastner,girshprivate} The position of
the peak in $YBCO$
has been reexamined in recent studies~
\protect\onlinecite{girshprivate}, which place the maximum at
$\omega_i \sim 3~eV$ rather than $2.8 eV$, as in the figure.}
\label{peak}
\end{figure}

We now analyze how these results are related to the experimental
measurements in Fig.~\ref{twom} and Fig.~\ref{peak}.
In the Introduction we listed the key experimental features that
required explanation.
Here we list them again and comment on each of them.

In Fig.~\ref{twom}:

\noindent
(a) {\it Asymmetry of the two-magnon peak profile\/}: our theory predicts that
for $\omega$ smaller than $\omega^{res}_i$ the two-magnon peak profile
should be asymmetric with a ``shoulder-like'' behavior at
frequencies close to $\omega =4J$, due to  the triple resonance in
the Raman vertex.
This is consistent with experimental observations. In particular,
the experimentally measured two-magnon profile in $Pr_2CuO_4$ was
analyzed~\cite{tomenoetal} and found
to contain two peaks, a two-magnon peak at $3000~cm^{-1}$,
and a smaller one at $4000~cm^{-1}$, which is precisely as expected from our
calculations. Notice, however, that
the calculation of the relative intensity of the peaks is beyond the scope of
the present approach.

\noindent
(b) {\it Selection rules\/}:
The leading diagram in the resonance regime contributes to scattering
in both $B_{1g}$ and $A_{1g}$ geometries.
The signals in both geometries have been observed in the experiments.
Recall that the Loudon-Fleury theory predicts scattering only in the
$B_{1g}$ geometry.

\noindent
(c) {\it Stability of the two-magnon peak profile\/}:
Away from the resonance, the stability of the two-magnon peak profile
is consistent with our theory, as the enhancement of the Raman vertex
for $ 2< \lambda_i <3.5$ occurs near $\omega =4J$.
As $\omega_i$ approaches $\omega^{res}_i$, the situation becomes more
complex, as one cannot separately consider the peak due to magnon-magnon
interaction and the enhancement of the Raman vertex due to the triple
resonance.
The experimental data seem to indicate that the two-magnon peak
profile changes little as $\omega_i$ sweeps through the resonance.
This is not entirely consistent with our scenario, as the triple
resonance region shifts to lower magnon energies $\omega$ as
the photon frequency $\omega_i$ approaches $\omega^{res}_i$.
However, the magnon profile itself in the resonant region can only
be obtained by solving coupled equations for the enhancements due to
the triple resonance and the final state magnon-magnon interactions.
We have not yet been able to accomplish that, and we leave it for future
work. We regard this as the most important unresolved issue of all
that remain.
Notice also that the observed
distortion of the Raman profile at higher photon frequencies
 is qualitatively consistent with our main conclusion that
once the photon frequency passes the
top of the coherent quasiparticle band, one must consider a
 qualitatively different set of electronic states.
Our speculation at present is that at such frequences the dominant contribution
to Raman scattering comes from
an incoherent,
diffusive hole motion, which can more easily couple to multi-magnon
excitations.

In Fig.~\ref{peak}:

\noindent
(a) {\it A single peak\/}: Our theory predicts a {\it single\/} maximum in
the two-magnon peak intensity measured as a function of the incident
photon frequency.
On the contrary, on the basis of general considerations in Raman
scattering,\cite{cardonareview} we might have expected two peaks,
one at $\omega_i = 2\Delta$, and the other at $\omega_f =2\Delta$.
These are the so-called incoming and outgoing resonances.
In the more traditional area of phonon Raman scattering in
semiconductors,\cite{cardonareview,martinfalikovreview,kleinreview}
the two peaks are not always separately resolved. However, given
the large magnon energy scale in the cuprates, we might have
expected them to be resolvable.
An additional bit of evidence that triple resonance is a plausible
explanation is that, experimentally, the two-magnon peak is seen in
the interval of incoming photon frequencies about $4000~cm^{-1}$ in
width,
which for $J \sim 1000~cm^{-1}$ is of the same width as the interval
$2< \lambda_i<4$ in our theory, where the triple resonance is possible
for all directions of the magnon momenta $\bf q$.

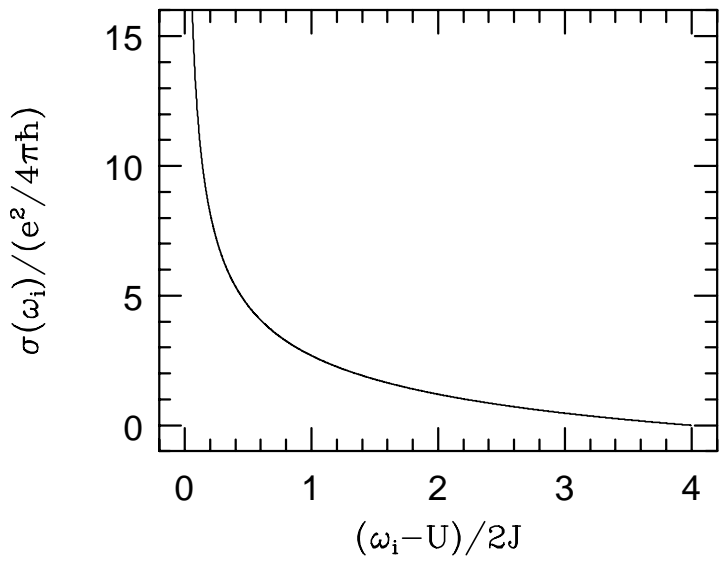
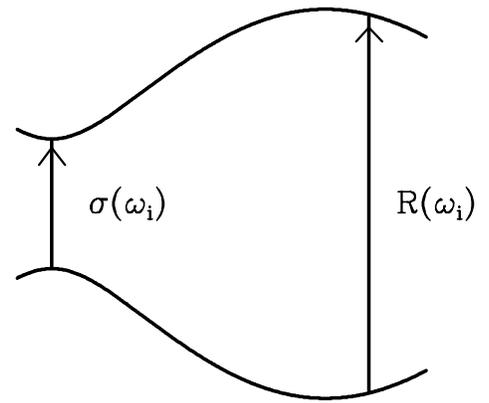
\begin{figure}
\caption{(a) The plot of Eq.~(\protect\ref{fullcond}) for the optical
conductivity, in the limit $t\ll U$.
 It has a square root divergence at the lower band
edge, and vanishes at the upper band edge. The divergence will be
washed out in practice, but we can still expect most prominent
features to be located towards the bottom of the SDW quasiparticle band.
(b) A schematic illustration of the relative location of
the strongest Raman and optical features as a function of the incident
photon frequency.
Their relative position is a key result of this work.
The two solid lines represent the fermion dispersion curves in the
conduction and valence bands. }
\label{newfig}
\end{figure}

\noindent
(b) {\it Peak location\/}:
Our theory predicts the  maximum of the two-magnon peak intensity,
measured as a function of $\omega_i$, right near the upper edge of
the quasiparticle fermionic band.
At the same time, the coherent quasiparticle contribution to the
optical conductivity $\sigma(\omega)$ was shown to have most of
its weight located near the lower band edge and vanish at the upper band edge
(as schematically illustrated in Fig.~\ref{newfig}(b)).
This is consistent with the experimental Fig.~\ref{peak}
which shows that Raman scattering is strongest right at
the upper end of those features in the optical data that can be
interpreted as particle-hole excitations between the lower and
upper Hubbard bands.
The location of the Raman maximum versus the features in
$\sigma(\omega)$ has been one of the greatest experimental
puzzles in this area, and its solution in our approach serves
as a partial verification of the SDW dispersion relation for the
carriers which, despite much theoretical work, has not been
well-established experimentally in these materials.

In Fig.~\ref{fit} we fit the published
experimental data~\cite{ranliu} from Fig.~\ref{peak}
on the peak intensity
in $YB_2Cu_3O_{6}$ at room temperature by our Eq.~(\ref{res}).
We see that a fit to the predicted inverse linear dependence is reasonably
good. The inverse linear dependence starts from $\omega_i \sim 2.4~eV$ and
extends
nearly up to the resonance frequency $\omega^{res}_i \approx 3.1~eV$.
The effects of fermionic damping are probably
relevant only
in the immediate vicinity of the resonance.
Note that the resonance value obtained from the fit
is larger than the peak location at $\omega^{res}_i \sim 2.8~eV$
in Fig~\ref{peak}.
In fact,  the experimental data at high frequencies have been
reexamined in recent studies~\cite{girshprivate}.
It turns out that the resonance at room temperature
occurs at a  higher frequency than
has been previously believed, and in particular the
the intensity at $\omega_i \sim 2.9~eV$ is found to be {\it larger}
than at $\omega_i \sim 2.8 ~eV$, unlike what one sees
in Fig.~\ref{peak}.
The new data suggest the resonance frequency value may exceed
$3~eV$, which would make it even more consistent with the fit to
the theoretical dependence.

These recent experiments~\cite{girshprivate} also measured the peak intensity
in $YB_2Cu_3O_{6.1}$ at low temperatures ($T\sim 5 K$).
We fitted the new data by the same inverse linear dependence and found
even better agreement with the experimental data for $\omega_i$ between
$2.5~eV$ and $3~eV$.  This fit yields about the same
 resonance frequency  $\omega^{res}_i \sim 3.1~eV$ as the fit at room
temperatures. We actually do not expect that the resonance values at
room temperatures and at $5 K$ are the same, but we notice that
the optical studies in $La_2CuO_4$ have shown\cite{kastner}
that the optical absorption peak shifts to higher frequencies by less
than $0.1~eV$ between room temperatures and $122~K$, and
one can expect the  peak in Raman intensity to shift by about the
same amount. This small shift is indeed within the combined accuracy
of our theory and the data available for the fits.

Notice that our theory also predicts that the Raman intensity should have
an additional, smaller, peak at $\omega \sim 2\Delta \sim 1.7 ~eV$
due to the Loudon-Fleury
mechanism. There is no
evidence for the peak at $\omega_i = 2\Delta$ in the room temperature data
on Fig~\ref{peak}.
Preliminary results of recent
low-temperature experiments~\cite{girshprivate2}
 indicate a possibility that there is in fact
a second, smaller peak
at $\omega_i \sim 1.6~eV$~which is consistent with our prediction.

There are, however, several experimental results which are beyond the scope of
our approach.
First  is the width of the two-magnon peak, which exceeds the prediction
of the spin-wave theory even if we take into account the resonance in
the Raman vertex.
As we already remarked above, simply invoking quantum fluctuations
for spin-$1/2$ is unlikely to improve the spin-wave theory, since
direct numerical calculations of the Raman spectrum on finite clusters
also predict a narrow peak in this case.\cite{norigalianobacci}
Another experimental fact is the existence of a  considerable Raman
signal $R(\omega)$ above the maximum possible two-magnon energy
(i.e., above $4J_{eff}$).
Canali and Girvin\cite{canaligirvin} performed a very detailed study
of the effects of four-magnon scattering within the context of
the Heisenberg model for spins and the Loudon-Fleury coupling of spins
to light.
They found that this scattering can give rise to the Raman cross-section
above $4J_{eff}$, but the intensity of the Raman signal was found to be too
small to fully account for the experimental data.
Thus one apparently has to go beyond the minimal model.

In the approach put forward by Singh,\cite{singhreview,singhetal}
further neighbor terms in the
effective Hamiltonian for coupling of light to spins are postulated,
and the coefficients adjusted so that the moments of the Raman
spectrum evaluated by the series expansion agree with observations.
Such next-nearest-neighbor terms broaden the two-magnon peak and also
automatically break the selection rule for the Loudon-Fleury
Hamiltonian which disallows scattering in the $A_{1g}$ configuration.
Of course, a question then arises of the origin of such terms and their
relative strength.
Notice that the further-neighbor terms are in fact also effectively
present in our momentum-space description, as it clearly follows from
the Shraiman-Shastry formulation of the problem in which the
non-Loudon-Fleury terms in the Raman vertex correspond to the effective
spin interactions between further neighbors.

In the bulk of the paper we ``factorized'' the problem by setting the
Raman vertex to its Loudon-Fleury form in the calculations of
two-magnon profile right near the peak.
This, however, is only an approximation, and in view of Singh's
results, we can expect some broadening of the two-magnon peak in
the more sophisticated calculations along our lines, which would
treat the resonance and the final state interactions together.
Simultaneously, we might also expect a shift of the two-magnon
peak position from $2.8J_{eff}$.
In the bulk of the paper we used $2.8J_{eff}$ because of the agreement with
neutron experiments.
We note however, that the only observations to date of the two-magnon
peak in cuprates were made at or near the resonant frequencies,
which are in the visible light region in cuprates.
In that sense, we may not even know what the "true", i.e., nonresonant
peak shape and location are.

The scattering in $B_{1g}$ and $A_{1g}$ geometries  in the SDW technique
has been studied by Kampf and Brenig.\cite{kampfbrenig}
They, however, focused on  high {\it transferred\/} frequencies, comparable
to the Mott-Hubbard gap, and did not include collective spin fluctuation
modes (i.e., magnons) in their theory. In view of this, a comparison of
our results with theirs is not possible.

In a rather different spirit, it was observed in a paper by Weber and
Ford\cite{weberford}
that even a rather small magnon damping introduced phenomenologically into
the equations broaden's the Parkinson peak considerably.
More recently, Merlin\cite{merlinpreprint1}
argued that phonons may be the source of that damping.
Should the electron-phonon interaction prove relevant to the problem,
this will only lend credence
of the key phenomenological assumption of this work, namely that on a
``first pass'' at the resonant scattering problem, it is best not
to try to deal simultaneously with two quite possibly distinct issues of
the two-magnon peak shape and its strength variation with the incoming
photon frequency.

Finally, an experimental fact which needs answering is the existence
of the Raman signal at {\it very} high incident and transferred frequencies,
of the order of $3.5~eV$ and $1~eV$, respectively.
The largest intensity at such high frequencies was
obtained in the ``chiral'' $A_{2g}$ channel.  Khveshchenko and
Wiegmann recently performed an  elegant study of the contribution to Raman
vertex in this frequency range
from chiral spin fluctuations in the magnetically ordered
phase.\cite{pasha}
For our considerations, it is essential that at such high frequencies,
the incoming photon is above
the coherent quasiparticle band (where our theory applies) and the motion
of the carriers that subsequently couple to the magnons is itself
incoherent in character.
This and other interesting issues require further experimental data
as well as an improved theoretical understanding.

\section{acknowledgements}
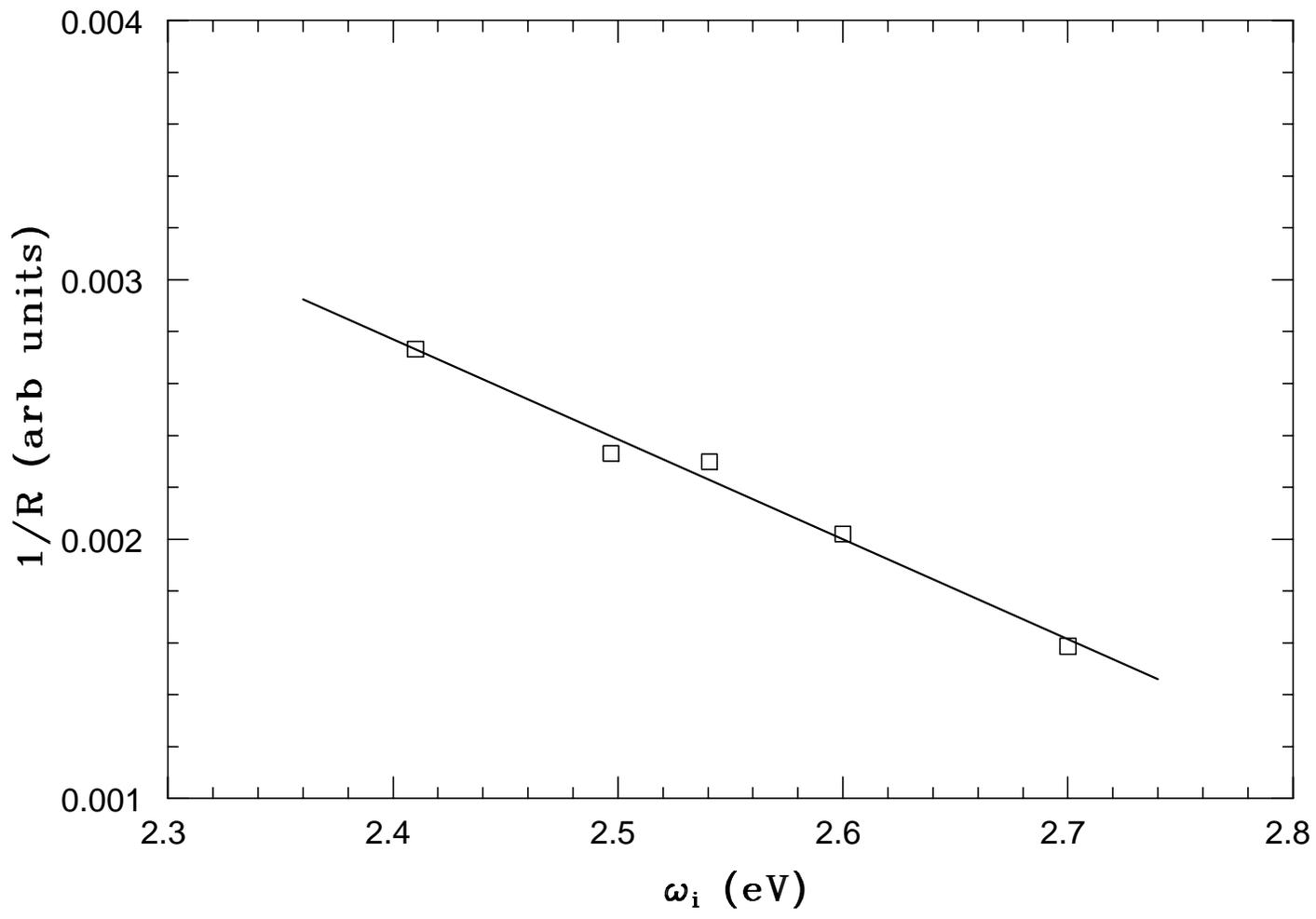
\begin{figure}
\caption{A fit of the experimental
dependence of the inverse two-magnon peak intensity in $YB_2Cu_3O_{6}$
from Fig.~\protect\ref{peak} to the theoretical
$1/(\omega^{res}_i-\omega_i)$ dependence,
Eq.~(\protect\ref{res}).
 The value of $\omega^{res}_i$ from the fit is $\approx 3.1 eV$.}
\label{fit}
\end{figure}

It is our pleasure to thank G.~Blumberg, C.~Canali, S.~L.~Cooper,
D.~Khveshchenko, M.~V.~Klein, R.~Liu, R.~Martin, R.~Merlin, H.~Monien,
D.~Pines, S.~Sachdev, C.~M. Varma, P.~Wiegmann, and A.~Zawadowski
for useful discussions and comments. We are grateful to G. Blumberg,~
M.~V.~Klein, S.~L. Cooper
and R.~Liu for providing experimental data for our Figs.~\ref{twom},
{}~\ref{peak}, and \ref{fit}.
D.F. is supported by the Texas Center for Superconductivity at
the University of Houston.
Part of the work has been done while A.C. was at Yale University,
where he was supported by the National Science Foundation (NSF)
Grant Nos. DMR-8857228 and
DMR-9224290, and while D.~F. was at the University of Illinois,
where he was supported by the NSF (DMR 91-20000) through the
Science and Technology Center for Superconductivity at the
University of Illinois at Urbana-Champaign.

\section{Appendix}

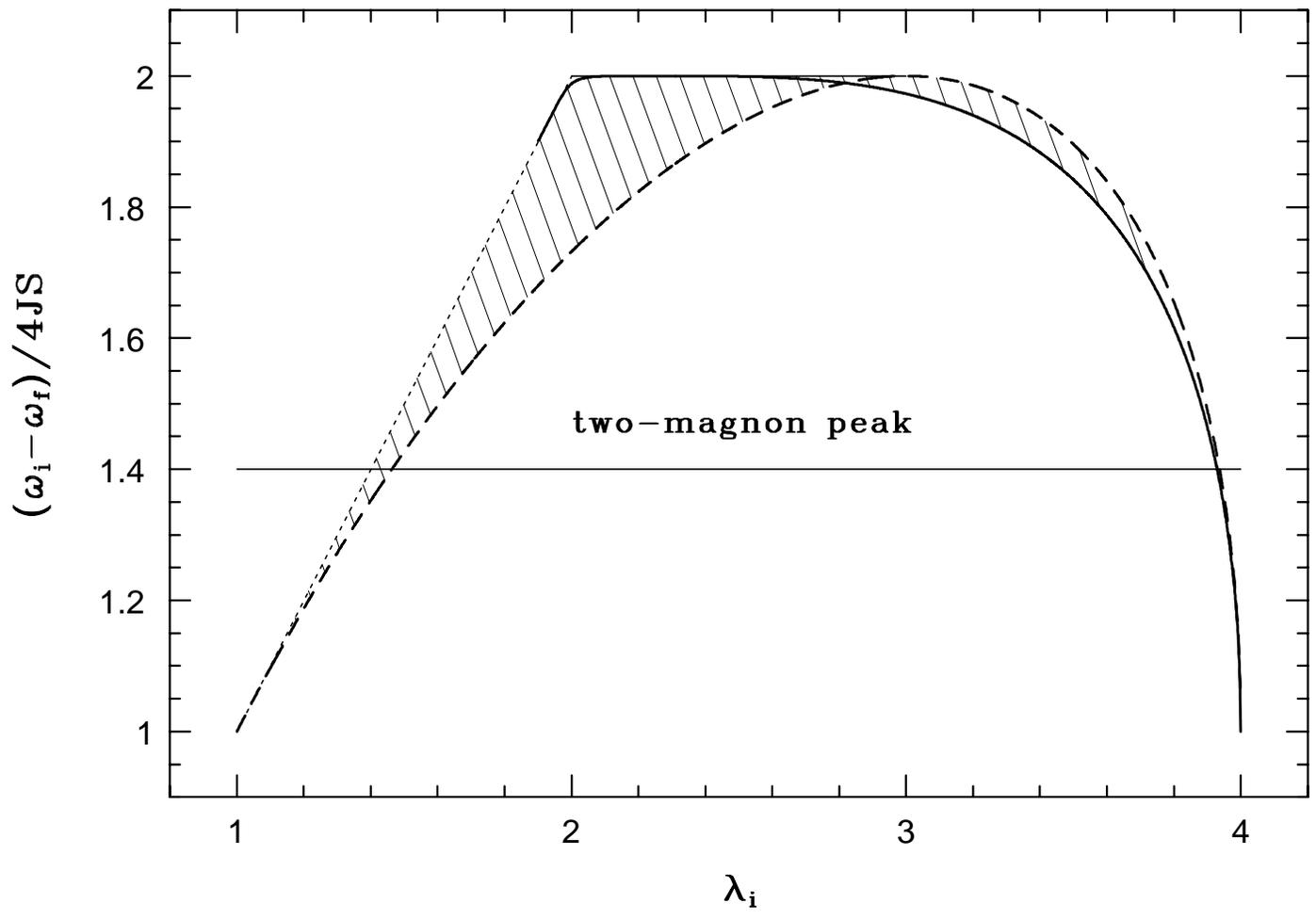
\begin{figure}
\caption{The triple resonance region (shaded) in the
$(\lambda_i,~(\omega_i - \omega_f)/4JS)$ plane where
$\lambda_i =(\omega_i -2\Delta)/4JS$.
Dashed line - solution for a triple resonance for magnon momentum $q_x = q_y$;
solid line - solution for $q_x =0$ or $q_y =0$.
The horisontal line corresponds to the position of the two-magnon peak at
$(\omega_i - \omega_f)/4JS = 2{\bar\omega} = 1.4$.}
\label{solution}
\end{figure}

In this Appendix, we obtain Eq.~(\ref{si}) for the Raman matrix element
near the intersection between $\omega = 2.8J_{eff}$, where the two-magnon peak
occurs, and a (narrow) region in the $(\lambda_i,\bar{\omega})$ plane
where triple resonance is allowed.  For simplicity, we will neglect the
difference between $J$ and $J_{eff}$.
We see from Fig.~\ref{solution} that this intersection occurs very close
to the top of the quasiparticle fermionic band
$E_k=(\Delta^2+ 4t^2(\cos k_x + \cos k_y)^2)^{1/2}$, i.e.,
near ${\bf k}={\bf 0}$.
Once this is established for this band structure, we can get a good
approximation for the integrals by expanding the energies to quadratic
order around ${\bf k}={\bf 0}$.

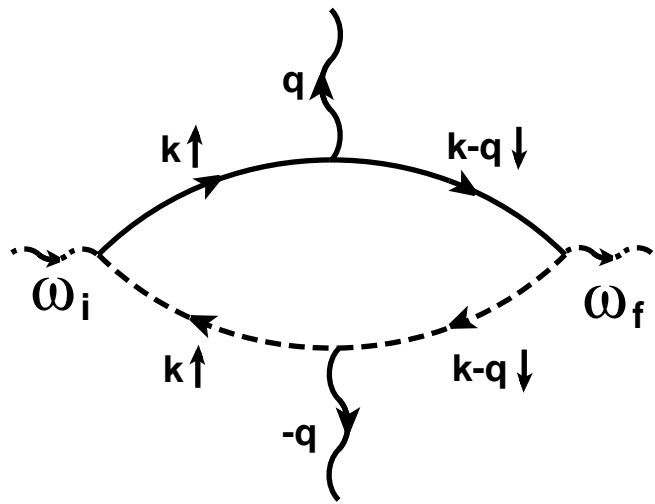
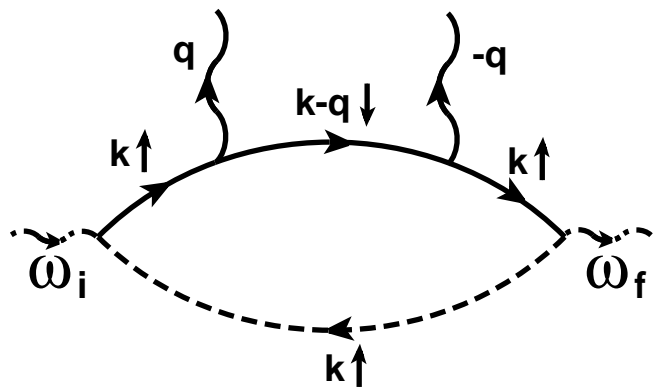
\begin{figure}
\caption{Diagrams which become important at resonance. All three
denominators in the diagram (a) can vanish simultaneously which is known as
a triple resonance. This diagram is dominant in the resonance region.
Additional relevant diagrams are generated by the spin label reversal
(in both (a) and (b)) and/or the emission of two magnons from the valence
band fermion line (in the diagram (b)).}
\label{intraband}
\end{figure}

The integral of interest is Eq.~(\ref{dia5}) for the diagram in
Fig.~\ref{intraband}(a):
\begin{equation}
M_R = -8i {\sum_k}'
{ \left( {\partial\epsilon_k\over\partial{\bf k}}\cdot{\bf\hat e}_i \right)
\left( {\partial\epsilon_{k-q}\over\partial{\bf k}}\cdot{\bf\hat e}^*_f
\right)
\left[ \mu_q\epsilon_{k-q}-\lambda_q\epsilon_k \right]^2
\over
\left( \omega_i-2E_k+i\delta \right)
\left( \omega_f-2E_{k-q}+i\delta \right)
\left( \omega_i-\Omega_q-E_k-E_{k-q}+i\delta \right) }
  \label{add13}
  \end{equation}

We have upon expanding
  \begin{equation}
\epsilon_k=
-2t (\cos k_x + \cos k_y)
\simeq
-4t \left(1-{{\bf k}^2 \over 4}\right),
  \label{add14}
  \end{equation}
and (we consider $S=1/2$)
  \begin{equation}
E_k =
\sqrt{\Delta^2+\epsilon_k^2}
\simeq
\Delta +4J\left(1-{{\bf k}^2 \over 2}\right).
  \label{A1}
  \end{equation}
We will expand both $E_k$ and $E_{k+q}$, which implies that we also assume
that the typical magnon momenta are not too large.
This, as we discuss in the bulk of the paper, is consistent with our
explicit result that the width of the region where triple resonance
is allowed is very narrow, nearly a single line.
To the same accuracy we can also linearize the magnon spectrum, $\Omega_q =
\sqrt{2} J q$.
However, we found it convenient to keep using the general form
$\Omega_q$ for the magnon frequency in some of the formulae.

To study the singular behavior, we first set the numerator to a constant.
Expanding in powers  of $\bf k$ in (\ref{A1}) and introducing, as before, the
reduced variables, $\lambda_i=(\omega_i-2\Delta)/2J$,
$\lambda_f=(\omega_f-2\Delta/)2J$, and ${\bar\Omega}_q = \Omega_q/2J$,
we obtain the following integral,
  \begin{equation}
I=\int d^2 k
{
1 \over
\left( \lambda_i-4+2{\bf k}^2+i\delta \right)
\left( \lambda_f-4+2({\bf k}-{\bf q})^2+i\delta \right)
\left( \lambda_i-{\bar\Omega}_q-4+{\bf k}^2 + ({\bf k}-{\bf q})^2+i\delta
\right)
}.
  \label{add15}
  \end{equation}

Somewhat similar integrals appear in the three-dimensional problem of
two-phonon resonant Raman scattering, where the incoming photon
frequency is tuned to match the critical points in the
semiconductor band structure.
The case when the hole band has an infinite mass was done analytically
by R.~Martin.\cite{martin}
The case of equal effective masses in two dimensions has not, to our
knowledge, been studied analytically before, and we present the
derivation in some detail.

We first observe that one of the terms in the denominator of the
integrand is half the sum of two others.
The integrand  can thus be put in the form
  \begin{equation}
{1\over AB \left( {A+B\over 2} \right) }
=
{1 \over 2 \left( {A+B\over 2} \right)^2 }
\left( {1\over A} + {1\over B}\right).
  \label{add16}
  \end{equation}

We now recall Feynman identity familiar from Quantum Electrodynamics,
  \begin{equation}
{1\over P^2Q}=\int^1_0 dx {2x\over(xP+(1-x)Q)^3}.
  \label{add17}
  \end{equation}
Using this identity and completing the squares containing $\bf k$, as
necessary, we obtain
  \begin{eqnarray}
I=\int d^2k \int_0^1 dx \Bigg[&&
{x\over
\left(\lambda_i-4+i\delta-x {\bar\Omega}_q-({x^2\over 2}-x)q^2
+2({\bf k}-{x\over2}{\bf q})^2 \right)^3 } \nonumber \\
+&&
{x\over
\left(\lambda_i-4+i\delta-(2-x) {\bar\Omega}_q-({x^2\over 2}-x)q^2
+2({\bf k}-(1-{x\over2}){\bf q})^2 \right)^3 }
\Bigg].
  \label{add18}
  \end{eqnarray}

Shifting the variables of integration and using  the integral
  \begin{equation}
\int d^2k {1\over (S+2k^2)^3}={\pi\over4S^2},
  \label{add19}
  \end{equation}
we obtain
  \begin{equation}
I={\pi\over4}\int_0^1 dx
\left[
{x\over\left(\lambda_i-4+i\delta -x {\bar\Omega}_q
-({x^2\over2}-x)q^2\right)^2}
+
{x\over\left(\lambda_i-4+i\delta -(2-x){\bar\Omega}_q
-({x^2\over2}-x)q^2\right)^2}
\right].
  \label{A2}
  \end{equation}

Completing further
the squares with respect to the variable of integration $x$,
we rewrite the r.h.s. of (\ref{A2}) as $I = I_1 + I_2$, where
  \begin{eqnarray}
I_1&=& {\pi\over4}\int_0^1 dx
{x \over
\left[\lambda_i-4+ {q^2\over2}\left(1-{{\bar\Omega}_q\over q^2}\right)^2+
i\delta - {q^2\over2}\left( x- (1-{{\bar\Omega}_q\over q^2})\right)^2 \right]^2
},
\nonumber \\
I_2 &=& {\pi\over4}\int_0^1 dx {x \over
\left[\lambda_i-4+ {q^2\over2}\left(1-{{\bar\Omega}_q\over q^2}\right)^2+
i\delta - {q^2\over2}\left( x- (1+{{\bar\Omega}_q\over q^2})\right)^2 \right]^2
}.
  \label{add20}
  \end{eqnarray}

At this stage, we have reduced $I$ to two integrals of the general type
  \begin{equation}
J=\int_0^1 dx{x\over\left[C-F^2(x-D)^2\right]^2}.
  \label{add21}
  \end{equation}
In both $I_1$ and $I_2$, we have the same
  \begin{equation}
C=\lambda_i-4+{q^2\over2}\left(1-{{\bar\Omega}_q\over q^2}\right)^2+i\delta
  \label{add22}
  \end{equation}
and $F^2={q^2\over2}$. However, the $D$'s are not the same. We have
  \begin{equation}
D_1=\left(1-{{\bar\Omega}_q\over q^2}\right),
  \label{add23}
  \end{equation}
and
  \begin{equation}
D_2=\left(1+{{\bar\Omega}_q\over q^2}\right)
  \label{add24}
  \end{equation}
in $I_1$ and $I_2$, respectively.

We now rewrite $J$ as
  \begin{equation}
J=
\int_0^1 dx{x-D\over\left[C-F^2(x-D)^2\right]^2}
 -{\partial\over\partial C}
\int_0^1 dx{D\over C-F^2(x-D)^2}.
  \label{add25}
  \end{equation}
Performing the integrations and partial differentiation with respect
to parameter, we obtain
  \begin{equation}
{J=
{1\over 2F^2}{1\over \left( C-F^2(x-D)^2 \right) }\bigg|^1_0 +
{D\over 2C} {x-D\over\left( C-F^2(x-D)^2\right)} \bigg|^1_0 +
{D\over 4F}{1\over C^{3\over2}}
\log\left({ F(x-D)+\sqrt{C}\over F(x-D)-\sqrt{C} }\right) \bigg|^1_0}.
  \label{add26}
  \end{equation}

The integral $I$ is a sum of two such integrals and is thus
  \begin{eqnarray}
I=I_1 + I_2 ={\pi\over 4}\Bigg\{&&
{1\over 2F^2}{1\over \left( C-F^2(x-D_1)^2 \right) }\bigg|^1_0
+ {1\over 2F^2}{1\over \left( C-F^2(x-D_2)^2 \right) }\bigg|^1_0
\nonumber \\
+&& {D_1\over 2C} {x-D_1\over\left( C-F^2(x-D_1)^2\right)} \bigg|^1_0
+ {D_2\over 2C} {x-D_2\over\left( C-F^2(x-D_2)^2\right)} \bigg|^1_0
\nonumber \\
+&& {D_1\over 4F}{1\over C^{3\over2}}
\log\left({ F(x-D_1)+\sqrt{C}\over F(x-D_1)-\sqrt{C} }\right) \bigg|^1_0
+ {D_2\over 4F}{1\over C^{3\over2}}
\log\left({ F(x-D_2)+\sqrt{C}\over F(x-D_2)-\sqrt{C} }\right) \bigg|^1_0
\Bigg\}.
  \label{add27}
  \end{eqnarray}

It turns out that the first four terms inside the braces add up to
zero. To see this, we
observe that the  definitions of $D_1$ and $D_2$ given above imply that
  \begin{equation}
(1-D_1)=-(1-D_2),
  \label{add28}
  \end{equation}
which in turn implies that $(1-D_1)^2=(1-D_2)^2$, and
  \begin{equation}
{D_1\over 2(1-D_1)} + {D_2\over 2(1-D_2)}+1=0.
  \label{add29}
  \end{equation}
We also use the identity
  \begin{equation}
{1\over C}{1\over (C-P)}=\left({1\over C-P} -{1\over C}\right )
{1\over P}
  \label{add30}
  \end{equation}
whenever terms of this type occur.
Using all these identities, we can easily show by a direct calculation
the cancellation claimed above.

We now have a compact result
  \begin{equation}
I={\pi\over4}
{1\over C^{3\over2}}
\Bigg\{
 {D_1\over 4F}
\log\left({ F(x-D_1)+\sqrt{C}\over F(x-D_1)-\sqrt{C} }\right) \bigg|^1_0
+ {D_2\over 4F}
\log\left({ F(x-D_2)+\sqrt{C}\over F(x-D_2)-\sqrt{C} }\right) \bigg|^1_0
\Bigg\}.
  \label{add31}
  \end{equation}

Substituting the full expressions for , $D_1$, $D_2$, and $F$, we obtain
  \begin{eqnarray}
I={\pi\over4}{1\over C^{3\over2}}  \Bigg\{&&
{\left(1-{{\bar\Omega}_q\over q^2}\right)\over 2\sqrt{2}q} \left[
\log\left({1+{\sqrt{2}q\over{\bar\Omega}_q}\sqrt{C}\over
           1-{\sqrt{2}q\over{\bar\Omega}_q}\sqrt{C}}\right)-
\log\left({1-{\sqrt{2}\over q\left(1-{{\bar\Omega}_q\over
q^2}\right)}\sqrt{C}\over
           1+{\sqrt{2}\over q\left(1-{{\bar\Omega}_q\over
q^2}\right)}\sqrt{C}}\right)
-2\pi i \right] \nonumber \\
+&&
{\left(1+{{\bar\Omega}_q\over q^2}\right)\over 2\sqrt{2}q} \left[
\log\left({1-{\sqrt{2}q\over{\bar\Omega}_q}\sqrt{C}\over
           1+{\sqrt{2}q\over{\bar\Omega}_q}\sqrt{C}}\right)-
\log\left({1-{\sqrt{2}\over q\left(1+{{\bar\Omega}_q\over
q^2}\right)}\sqrt{C}\over
           1+{\sqrt{2}\over q\left(1+{{\bar\Omega}_q\over
q^2}\right)}\sqrt{C}}\right)
\right]
\Bigg\}.
  \label{A3}
  \end{eqnarray}
The vanishing of $C$
corresponds to the triple resonance. Near this point,
$I$ behaves as  $I \sim C^{-3\over 2}$.

The expression above is incomplete until we specify the branches
of the logarithms and the square roots.
A careful examination of the derivation shows that we must choose
the square root branch which is positive for $C > 0$ and the
branch of the logarithms which tends to zero when its argument goes
to unity. The answer for other real values of $C$ is obtained by
setting $C \to C+ i\delta$ and then analytically continuing $C$
along the real axis.

We now turn to the form-factor in the numerator of Eq.~(\ref{add13}).
Clearly, for any $\lambda_i^{res} <4$, this form-factor can be
considered constant in the immediate vicinity of $\lambda_i^{res}$.
At the same time, since the form-factor vanishes at the
top of the quasiparticle band, it will contribute a small factor to
Raman intensity for $\lambda_i^{res} \sim 4$. This will
effectively change the
singularity of the Raman matrix element at some distance away from
the triple resonance, since for $\lambda_i$ away from 4 the difference
between $\lambda_i^{res}$ being exactly 4 or merely close to it is not
noticeable.

Near $\lambda_i^{res}=4$,
we set $(\mu_q\epsilon_{k-q}-\lambda_q\epsilon_k)=const$,
linearize ${\partial\epsilon_k\over\partial{\bf k}}\sim {\bf k}$
and ${\partial\epsilon_{k-q}\over\partial{\bf k}}\sim {\bf k-q}$,
expand the numerator terms to quadratic order,
and obtain
  \begin{equation}
  M_R\sim {\sum_k}'
  {
     [{\bf k}\cdot{\bf\hat e}_i]
     [({\bf k-q})\cdot{\bf\hat e}^*_f]
  \over
\left( \lambda_i-4+2{\bf k}^2+i\delta \right)
\left( \lambda_f-4+2({\bf k}-{\bf q})^2+i\delta \right)
\left( \lambda_i-{\bar\Omega}_q-4+{\bf k}^2 + ({\bf k}-{\bf q})^2+i\delta
\right)
  }
  \label{formf1}
  \end{equation}
  in place of Eq.~(\ref{add15}) above.
  Then we proceed using Feynman parametric representation as in the case
  with a constant numerator above.
  The two shifted $\bf k$ variables in Eq.~(\ref{add18}) were
  ${\bf k'}={\bf k}-{x\over 2} {\bf q} $ and
  ${\bf k''}={\bf k}-\left(1-{x\over 2}\right){\bf q}$.
  We can alternately express the numerator in the integral above in terms
  of these as
  \begin{equation}
  [{\bf k}\cdot{\bf\hat e}_i]
  [({\bf k-q})\cdot{\bf\hat e}^*_f]=
  \left[\left({\bf k'}+{x\over 2}{\bf q}\right)
  \cdot{\bf\hat e}_i\right]
  \left[ \left({\bf k'}-\left(1-{x\over 2}\right){\bf q}\right)
  \cdot{\bf\hat e}^*_f\right],
  \label{formf2}
  \end{equation}
  and
  \begin{equation}
  [{\bf k}\cdot{\bf\hat e}_i]
  [({\bf k-q})\cdot{\bf\hat e}^*_f]=
  \left[\left({\bf k''}+\left(1-{x\over 2}\right){\bf q}\right)
  \cdot{\bf\hat e}_i\right]
  \left[\left({\bf k''}-{x\over 2}{\bf q}\right)
  \cdot{\bf\hat e}^*_f\right].
  \label{formf3}
  \end{equation}

  When we now integrate over $\bf k'$ and $\bf k''$ in the $B_{1g}$
  scattering geometry, the angular integrals will only leave the terms
  $-{x\over 2}(1-{x\over 2})({\bf q}\cdot{\bf\hat e}_i)
   ({\bf q}\cdot{\bf\hat e}^*_f)$ in the numerator.
  Thus, we obtain the same integrals as in Eq.~(\ref{add20}) above,
  except that instead of $x$ in the numerators we will have
  $-{x^2\over 2}(1-{x\over 2})$. The angular factor involving magnon
  momenta, $({\bf q}\cdot{\bf\hat e}_i)
  ({\bf q}\cdot{\bf\hat e}^*_f)$, when computed in the
  $B_{1g}$ geometry, is just an expansion of the Loudon-Fleury vertex
  $M_R^{B_1}$ of Eq.~(\ref{B1g}) to quadratic order in $\bf q$,
  which is consistent with the linearization of the
  form-factor done in the beginning of this calculation.

  The integrals over the parametric variable $x$ can now
  be done by elemetary means similar to those employed above
  for a constant numerator case.
  The resulting expressions are, however, too lengthy to be worth presenting
  here.
  At the same time, as we discussed above, what is of real interest to us
  is the modification of the singular behavior of $M_R$ at the values of
  $\lambda_i$ at which one can neglect the difference between
  $\lambda_i - \lambda_i^{res}$ and $\lambda_i-4$.
  A simplest way to proceed is to consider the case $\lambda_i^{res}=4$,
  when we also have ${\bar\Omega}_q={1\over 2}$.
  In that case $D_1=0$ in Eq.~(\ref{add23}), and $D_2=2$ in Eq.~(\ref{add24}).
  Thus, of the two integrals like the one of Eq.~(\ref{add21}) above, but
  now with $-{x^2\over 2}(1-{x\over 2})$ instead of $x$ in the numerator,
  only the one with $D_1$ is singular for small $C$.
  The singularity is, furthermore, picked up near $x=0$, so that we can
  also omit the factor $(1-{x\over 2})$ in the numerator.
  We end up with
  \begin{equation}
  M_R\sim\int_0^1{ x^2
  \over
  [4C- x^2]^2},
  \label{formf4}
  \end{equation}
  with $C=\lambda_i-4+i\delta$.

  Performing an elemetary integration, and omitting an irrelevant
  nonsingular term, we obtain
  \begin{equation}
  M_R\sim
  {2\over\sqrt{C}}
  \left(\ln\left({1-2\sqrt{C}\over 1+2\sqrt{C}}\right)+i\pi\right),
  \label{formf5}
  \end{equation}
  which has a leading square root singularity $2\pi i\over\sqrt{C}$.
  This is the result needed in the main text. Combining now Eqs.~(\ref{A3})
  and (\ref{formf5}) and restricting to only a leading singularity,
  we obtain the result quoted in Eq.~(\ref{si}) in the main text.

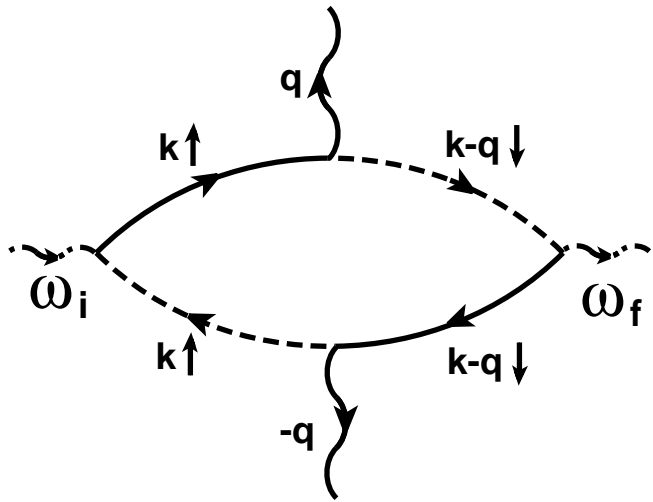
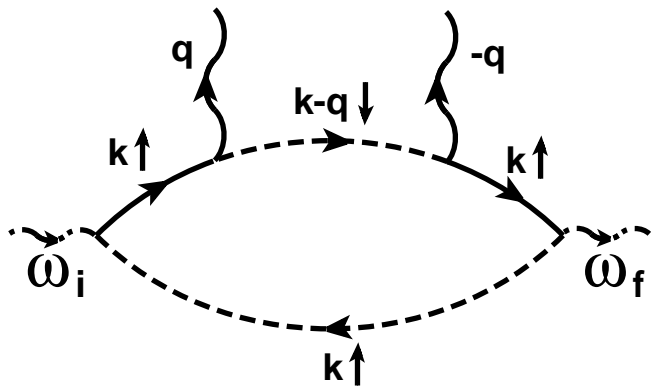
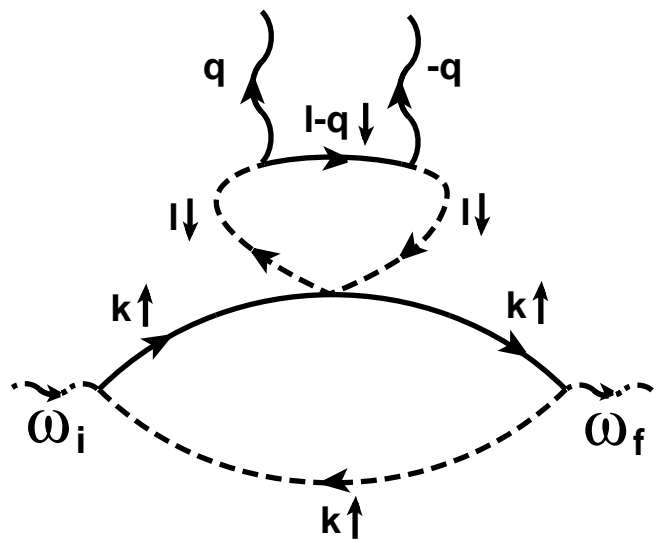
\begin{figure}
\caption{The diagrams for two-magnon emission which contribute to the
Loudon-Fleury Hamiltonian at small incident frequencies.
Each fermion can belong to either the valence (dashed line) or conduction
(solid line) band.
The emitted magnons are denoted by the solid wavy lines, and the incoming
($\omega_i$) and outgoing ($\omega_f$) photons by the dash-dotted
lines at the ends of the diagrams.
Additional graphs are obtained from each diagram by the fermion flow
reversal and/or flipping all the spin labels.}
\label{nonresdiag}
\end{figure}

\begin{references}
\bibitem{SScom} We  corrected a trivial factor of 4 in
Ref.~\onlinecite{shastryshraiman1}.
\bibitem{footnoteoncancellationsofsqarepieces}
Actually, each of the diagrams (a) and (b) in Fig.~\protect\ref{nonresdiag}
also has a term with a doubly resonant denominator.
However, the doubly resonant terms cancel when we add the two diagrams
(see Sec.~\protect\ref{ramanmatrixelement}).
\bibitem{comment} Notice that the fact that the second term is the
half-sum of the other two is the result of the particle-hole symmetry
in the Hubbard model at half-filling.
If it is violated, as may be the case in real materials, then,
strictly speaking, a triple resonance would not occur.
We will assume that the violation of particle-hole
symmetry can be neglected.
\bibitem{comm2} This is even more true
in the case of polar materials,
since the Frohlich electron-phonon interaction, which is
singular at the minimum of the fermionic band,
precludes large electron momentum changes, further confining the
electronic states to the vicinity of the critical point.
\bibitem{extracomm} Actually, for $\lambda_i <9/4$, there exists a
second solution for $q_x=0$, but this solution is located well inside the
shaded region in Fig~\protect\ref{solution} and is therefore irrelevant to
our considerations.
\end{references}
\end{document}